\documentclass[journal=jacsat,manuscript=article]{achemso}

\usepackage{chemformula} 
\usepackage[T1]{fontenc} 
\usepackage{eufrak}
\usepackage{bm}
\usepackage{amsmath,amssymb}

\newcommand{\parallelsum}{\mathbin{\!/\mkern-5mu/\!}}
\DeclareMathAlphabet\mathbfcal{OMS}{cmsy}{b}{n}

\def\ket#1{\mathinner{|{#1}\rangle}}

\newcommand{\onlinecite}[1]{\hspace{-1 ex} \nocite{#1}\citenum{#1}} 

\author{\small D. Hagenm\"{u}ller}
\email{dhagenmuller@unistra.fr}
\affiliation[ISIS]
{\small ISIS (UMR 7006) and icFRC, University of Strasbourg and CNRS, Strasbourg, France}
\author{\small J. Schachenmayer}
\affiliation[ISIS]
{\small ISIS (UMR 7006) and icFRC, University of Strasbourg and CNRS, Strasbourg, France}
\alsoaffiliation[IPCMS]
{\small IPCMS (UMR 7504), University of Strasbourg and CNRS, Strasbourg, France}
\author{\small C. Genet}
\affiliation[ISIS]
{\small ISIS (UMR 7006) and icFRC, University of Strasbourg and CNRS, Strasbourg, France}
\author{\small T. W. Ebbesen}
\affiliation[ISIS]
{\small ISIS (UMR 7006) and icFRC, University of Strasbourg and CNRS, Strasbourg, France}
\author{\small G. Pupillo}
\email{pupillo@unistra.fr}
\affiliation[ISIS]
{\small ISIS (UMR 7006) and icFRC, University of Strasbourg and CNRS, Strasbourg, France}
\alsoaffiliation[IPCMS]
{\small IPCMS (UMR 7504), University of Strasbourg and CNRS, Strasbourg, France}

\title[title]
  {Enhancement of the electron-phonon scattering induced by intrinsic surface plasmon-phonon polaritons}

\abbreviations{SPP,LP,UP}
\keywords{Plasmonics, Polaritons, Phonons, Plasmons, Ultrastrong coupling, Electron-phonon coupling, 2D materials}

\begin{document}


\begin{abstract}
We investigate light-matter coupling in metallic crystals where plasmons coexist with phonons exhibiting large oscillator strength. We demonstrate theoretically that this coexistence can lead to strong light-matter interactions without external resonators. When the frequencies of plasmons and phonons are comparable, hybridization of these collective matter modes occurs in the crystal. We show that the coupling of these modes to photonic degrees of freedom gives rise to intrinsic surface plasmon-phonon polaritons, which offer the unique possibility to control the phonon properties by tuning the electron density and the crystal thickness. In particular, dressed phonons with reduced frequency and large wave vectors arise in the case of quasi-2D crystals, which leads to large enhancements of the electron-phonon scattering in the vibrational ultrastrong coupling regime. This suggests that photons can play a key role in determining the quantum properties of certain materials. A non-perturbative self-consistent Hamiltonian method is presented that is valid for arbitrarily large coupling strengths.   
\end{abstract}

\textbf{Keywords:} Electron-phonon coupling, Polaritons, Phonons, Plasmons, Ultrastrong coupling, 2D materials

Using light to shape the properties of quantum materials is a long-standing goal in physics, which is still attracting much attention~\cite{schlawin,curtis,PhysRevLett.122.017401}. Since the 60s, it is known that superconductivity can be stimulated via an amplification of the gap induced by an external electromagnetic radiation~\cite{PhysRevLett.13.195,PhysRevLett.16.1166}. Ultrafast pump-probe techniques have been recently used to control the phases of materials such as magnetoresistive manganites~\cite{rini}, layered high-temperature superconductors~\cite{Fausti189}, as well as alkali-doped fullerenes~\cite{mitrano}, by tuning to a specific mid-infrared molecular vibration. It is an interesting question whether the phases of quantum materials may be also passively modified by strong light-matter interactions in the steady-state, without external radiation. Recently, the possibility of modifying superconductivity by coupling a vibrational mode to the vacuum field of a cavity-type structure has been suggested~\cite{thomas}. This idea was then investigated theoretically in the case of a ${\rm Fe}{\rm Se}/{\rm Sr}{\rm Ti}{\rm O}_{3}$ superconducting heterostructure embedded in a Fabry-Perot cavity~\cite{rubio}. In this work, enhanced superconductivity relies, however, on unrealistically large values of the phonon-photon coupling strength, in a regime where light and matter degrees of freedom totally decouple~\cite{PhysRevLett.112.016401}. This occurs beyond the ultrastrong coupling regime, which is defined when the light-matter coupling strength reaches a few tens of percents of the relevant transition frequency~\cite{ciuti1,Forn,Kockum}.

Thanks to the confinement of light below the diffraction limit~\cite{BarnesTW}, an alternative approach to engineer strong~\cite{AGRANOVICH1974169,YAKOVLEV1975293,doi:10.1063/1.443834_1982,PhysRevLett.93.036404,dintinger_2005,PhysRevLett.101.116801_2008,doi:10.1021/nl903455z_2010,doi:10.1021/nn201077r_2011,tudela_2013,orgiu2015conductivity,aizer_2017} and even ultrastrong light-matter coupling~\cite{Balci:13,doi:10.1021/nn504652w,doi:10.1021/acsphotonics.7b00554} is the use of plasmonic resonators. The latter allow the propagation of surface plasmon polaritons (SPPs), which are evanescent waves originating from the coupling between light and collective electronic modes in metals, called plasmons~\cite{raether1988surface}.  Interestingly, most quantized models involving SPPs are well suited to describe the strong coupling regime~\cite{hummer_2013,tudela_2013}, while a corresponding theory for the ultrastrong coupling regime is still missing. 

In contrast to SPPs, an Hamiltonian model valid for all interaction regimes of surface phonon polaritons stemming from the coupling between light and ionic charges already exists~\cite{PhysRevB.94.205301}. In the absence of external resonator, strong coupling between SPPs in a 2D material and surface phonons of the substrate has recently stimulated considerable interest~\cite{PhysRevB.82.201413,doi:10.1021/nl501096s,PhysRevLett.116.106802,Lin6717,doi:10.1021/acsphotonics.7b00928}. Such a situation can not generally occur in the same crystal due to the screening of the ion charges by electrons. Nevertheless, exceptions exist as a result of the coupling between phonons and interband transitions~\cite{PhysRevLett.39.1359_1977,PhysRevB.45.10173_1992}. These include bilayer graphene~\cite{PhysRevLett.103.116804_2009}, alkali fullerides ${\rm A}_{6} {\rm C}_{60}$~\cite{PhysRevB.46.1937_1992}, organic conductors (e.g. ${\rm K}-{\rm TCNQ}$~\cite{PhysRevB.16.3283_1977,doi:10.1063/1.437662_1979}), as well as some transition metal compounds~\cite{PhysRevB.55.R4863}. Interestingly, some of these materials or similar compounds feature superconductivity and charge density waves under certain conditions, which are believed to be driven, at least partly, by electron-phonon interactions~\cite{tani,supra_fullerides,doi:10.1021/cr030652g,yuan,PhysRevLett.121.257001}. It is an open question whether the ultrastrong coupling stemming from the coexistence of plasmons and phonons with large oscillator strengths can affect the electron-phonon scattering without external resonator.

In this work, we propose and investigate the possibility to tune material properties of certain crystals via an hybridization of phonons, plasmons, and photons which is intrinsic to the material, and without the use of an external resonator. The coexistence of plasmons and phonons within the same crystal gives rise to hybrid plasmon-phonon modes, which have been extensively studied in weakly-doped semiconductors and semimetals~\cite{mahan,PhysRevLett.46.500,PhysRevB.40.9723,PhysRevB.82.195406,doi:10.1021/acs.nanolett.7b01603}. Here, we show that while hybrid plasmon-phonon modes do not affect the electron-phonon scattering at the level of the random phase approximation, strong interactions between these modes and photons offer a unique possibility to control the energy and momentum of the resulting surface plasmon-phonon polaritons by tuning the electron density and the crystal thickness: ``Dressed'' phonons with reduced frequency arise in the ultrastrong coupling regime, where the electronic and ionic plasma frequencies become comparable to the phonon frequency. These dressed phonons are shown to exhibit unusually large momenta comparable to the Fermi wave vector in the case of quasi-2D crystals, which leads to large enhancements of the electron-phonon scattering. These results suggest that photons can play a key role in determining the quantum properties of certain materials.

We utilize a self-consistent Hamiltonian method based on a generalization of that introduced in Ref.~[\onlinecite{todorovY}], which is valid for all regimes of interactions including the ultrastrong coupling regime, and in the absence of dissipation. This quantum description of surface plasmon-phonon polaritons provides an ideal framework to investigate how the latter can affect the quantum properties of the crystal. Furthermore, our method can be generalized to various surrounding media and geometries of interest such as layered metallo-dielectric metamaterials and simple nano-structures, and differs from the usual effective quantum description of strong coupling between excitons and quantized SPPs~\cite{hummer_2013,tudela_2013}. We show in the supplemental material that the latter model leads to unphysical behaviors in the ultrastrong coupling regime.  

\vspace{3mm}

\noindent The paper is organized as follows: \textit{i)} We first introduce the total Hamiltonian of the system under consideration; and \textit{ii)} Diagonalize the matter part leading to plasmon-phonon hybridization in the crystal; \textit{iii)} We derive the coupling Hamiltonian of these hybrid modes to photonic degrees of freedom, which gives rise to intrinsic surface plasmon-phonon polaritons. \textit{iv)} We determine the Hamiltonian parameters and its eigenvalues self-consistently using the Helmoltz equation, and characterize the properties of the resulting surface plasmon-phonon polaritons; \textit{v)} We show that while plasmon-phonon hybridization can not solely modify the electron-phonon scattering at the level of the random phase approximation (RPA), the coupling of these hybrid modes to photons can lead to large enhancements of the electron-phonon scattering in the crystal.

\section{Quantum Hamiltonian}

\begin{figure}[ht]
\centerline{\includegraphics[width=0.9\columnwidth]{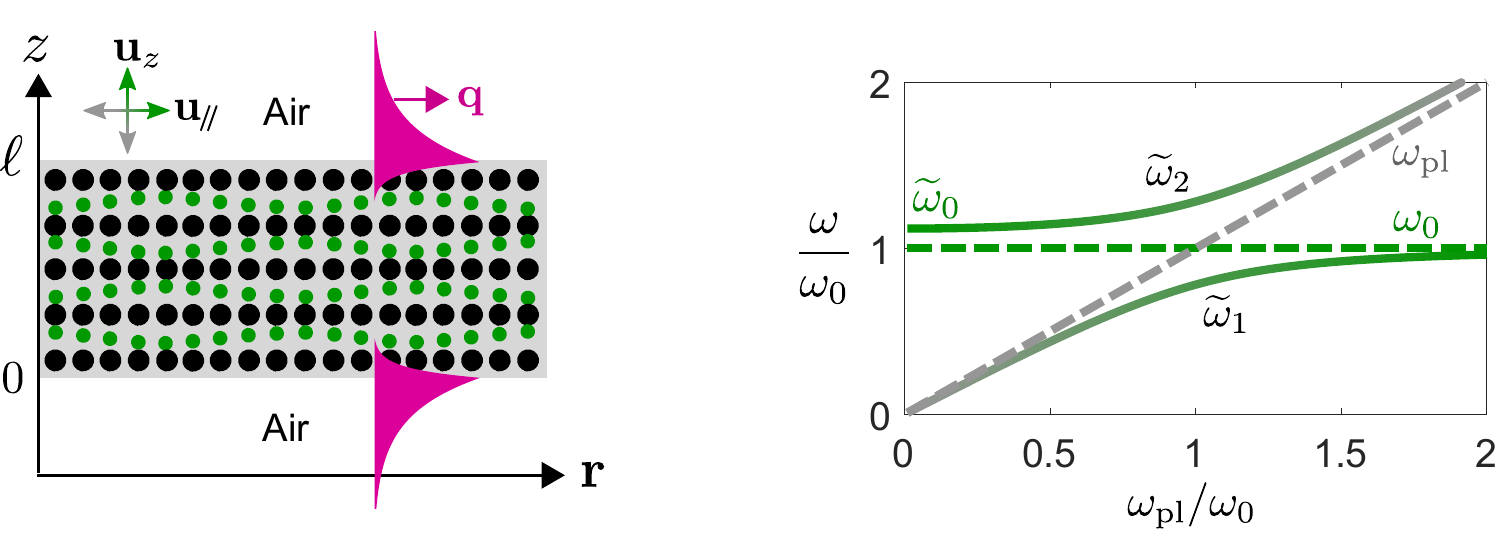}}
\caption{Intrinsic surface plasmon-phonon polaritons: Metal of thickness $\ell$ in air featuring a plasmon mode with frequency $\omega_{\rm pl}$ (gray dashed line) and an optical phonon mode with frequency $\omega_0$ (green dashed line). Both plasmons and phonons are polarized in the directions ${\bf u}_z$ and ${\bf u}_{\parallelsum}$. When $\nu_{\rm pn}\neq 0$ and $\omega_{\rm pl}\sim\omega_{0}$, the matter Hamiltonian $H_{\rm mat}$ provides two hybrid plasmon-phonon modes with frequencies $\widetilde{\omega}_{1}$ and $\widetilde{\omega}_{2}$ [Eq.~(\ref{polpol})] represented as a function of $\omega_{\rm pl}/\omega_{0}$ for $\nu_{\rm pn}=0.5\omega_{0}$. The surface polaritons generated by the coupling of these modes to light propagate along the two metal-dielectric interfaces with in-plane wave vector ${\bf q}$.} 
\label{fig1}
\end{figure}

We consider a metallic crystal of surface $S$ and thickness $\ell$ in air, which contains a free electron gas and a transverse optical phonon mode with frequency $\omega_{0}$. The excitation spectrum of the electron gas features a collective, long-wavelength plasmon mode with plasma frequency $\omega_{\rm pl}$, as well as a continuum of individual excitations called electronic dark modes that are orthogonal to the plasmon mode. Both plasmons and phonons are polarized in the directions ${\bf u}_z$ and ${\bf u}_{\parallelsum}$ (Fig.~\ref{fig1}). The Hamiltonian is derived in the Power-Zienau-Woolley representation~\cite{Babiker439}, which ensures proper inclusion of all photon-mediated dipole-dipole interactions, and can be decomposed as $H = H_{\rm pol} + H_{\rm el-pn}$ with $H_{\rm pol}=H_{\rm pt} +H_{\rm mat-pt}+ H_{\rm mat}$. The first term in $H_{\rm pol}$ reads $H_{\rm pt} = \frac{1}{2\epsilon_{0}}\int \!d{\bf R} \, {\bf D}^{2}({\bf R}) + \frac{1}{2\epsilon_{0}c^{2}}\int \!d{\bf R}\, {\bf H}^{2}({\bf R})$, and corresponds to the photon Hamiltonian. $c$ denotes the speed of light in vacuum, ${\bf R}\equiv ({\bf r},z)$ the 3D position, $\epsilon_{0}$ the vacuum permittivity, and ${\bf D}$ and ${\bf H}$ are the displacement and magnetic fields, respectively. The light-matter coupling term reads $H_{\rm mat-pt}=-\frac{1}{\epsilon_{0}}\int \!d{\bf R} \, {\bf P} ({\bf R}) \cdot {\bf D} ({\bf R})$. Here, ${\bf P}= {\bf P}_{\rm pl}+{\bf P}_{\rm pn}$ denotes the matter polarization field associated to the dipole moment density, where ${\bf P}_{\rm pl}$ and ${\bf P}_{\rm pn}$ correspond respectively to the plasmon and phonon contributions~\footnote{\label{footnote_1} Our method can be easily generalized when phonons are replaced by excitons, by considering the appropriate polarization field.}. The matter Hamiltonian is decomposed as $H_{\rm mat}=H_{\rm pn}+H_{\rm pl}+H_{\rm P^{2}}$, where $H_{\rm pn}$ and $H_{\rm pl}$ denote the contributions of free phonons and plasmons which are provided in the supplemental material, and $H_{\rm P^{2}}=\frac{1}{2\epsilon_{0}}\int \!d{\bf R} \, {\bf P}^{2} ({\bf R})$ contains terms $\propto {\bf P}^{2}_{\rm pl},{\bf P}^{2}_{\rm ph}$, and a direct plasmon-phonon interaction $\propto {\bf P}_{\rm pl}\cdot{\bf P}_{\rm ph}$. Finally, $H_{\rm el-pn}$ includes the contribution of the free, individual electrons (electronic dark modes), as well as the coupling Hamiltonian between electronic dark modes and phonons.

\section{Diagonalization of the matter Hamiltonian}

Hybrid plasmon-phonon modes have been extensively studied in weakly-doped semiconductors and semimetals~\cite{mahan,PhysRevLett.46.500,PhysRevB.40.9723,PhysRevB.82.195406,doi:10.1021/acs.nanolett.7b01603}. In this section, we propose a derivation of these hybrid modes using a non-perturbative quantum method. In order to diagonalize $H_{\rm mat}$, the phonon polarization ${\bf P}_{\rm pn}$ is written in terms of the bosonic phonon annihilation and creation operators $B_{{\bf Q},\alpha}$ and $B^{\dagger}_{{\bf Q},\alpha}$, with the 3D wave vector ${\bf Q} \equiv ({\bf q},q_{z})$ and $\alpha = \parallelsum, z$ the polarization index. The lattice polarization field ${\bf P}_{\rm pn}$ is proportional to the ion plasma frequency $\nu_{\rm pn}=\sqrt{\frac{\mathcal{Z}^{2}N}{M\epsilon_{0} a^{3}}}$, which plays the role of the phonon-photon coupling strength. Here, $N$ denotes the number of vibrating ions with effective mass $M$ and charge $\mathcal{Z}$ in a unit cell of volume $a^{3}$. The plasmon polarization ${\bf P}_{\rm pl}$ is provided by the dipolar description of the free electron gas corresponding to that of the RPA~\cite{todorovY}. In the long-wavelength regime, ${\bf P}_{\rm pl}$ is written in terms of the bosonic plasmon operators $P_{{\bf K}{\bf Q}}$ and $P^{\dagger}_{{\bf K}{\bf Q}}$ (${\bf K}$ denotes the 3D electron wave vector), superpositions of electron-hole excitations with wave vector ${\bf Q}$ across the Fermi surface. As explained in the supplemental material, $H_{\rm mat}$ can be put in the diagonal form $H_{\rm mat}= \sum_{{\bf Q},\alpha,j=1,2} \hbar \widetilde{\omega}_{j} \Pi^{\dagger}_{{\bf Q}j\alpha} \Pi_{{\bf Q}j\alpha}$, where the hybrid plasmon-phonon operators $\Pi_{{\bf Q}j\alpha}$ are superpositions of $B_{{\bf Q},\alpha}$, $P_{{\bf K}{\bf Q}}$, and their hermitian conjugates, and satisfy the bosonic commutation relations $[\Pi_{{\bf Q}j \alpha},\Pi^{\dagger}_{{\bf Q'}j' \alpha'}]=\delta_{{\bf Q},{\bf Q'}}\delta_{j,j'}\delta_{\alpha,\alpha'}$. The hybrid mode frequencies $\widetilde{\omega}_{j}$ are represented on Fig.~\ref{fig1} and are given by
\begin{align}
2\widetilde{\omega}^{2}_{j}=\widetilde{\omega}^{2}_{0} + \omega^{2}_{\rm pl} \pm \sqrt{\left(\widetilde{\omega}^{2}_{0} - \omega^{2}_{\rm pl} \right)^{2} + 4 \nu^{2}_{\rm pn} \omega^{2}_{\rm pl}}. 
\label{polpol}
\end{align} 
Here, $j=1$ and $j=2$ refer respectively to the signs $-$ and $+$, and the longitudinal phonon mode with frequency $\widetilde{\omega}_{0}=\sqrt{\omega^{2}_{0}+ \nu^{2}_{\rm pn}}$ is determined by the combination of the short-range restoring force related to the transverse phonon resonance, and the long-range Coulomb force associated to the ion plasma frequency. The diagonalization procedure of $H_{\rm mat}$ further provides the well-known transverse dielectric function of the crystal~\cite{mahan}:
\begin{align}
\epsilon_{\rm cr} (\omega)= 1 - \frac{\omega^{2}_{\rm pl}}{\omega^{2}}- \frac{\nu^{2}_{\rm pn}}{\omega^{2}-\omega^{2}_{0}}.
\label{epsi_cr}
\end{align} 
One can then use the eigenmodes basis of the matter Hamiltonian to express the total polarization field as
\begin{align}
{\bf P} ({\bf R})= \sum_{{\bf Q},\alpha,j} g_{j}\sqrt{\frac{\hbar \epsilon_{0} \omega^{2}_{\rm pl}}{2 \widetilde{\omega}_{j} V}} \left(\Pi_{{\bf -Q}j\alpha} + \Pi^{\dagger}_{{\bf Q}j\alpha}\right) e^{-i {\bf Q}\cdot {\bf R}} {\bf u}_{\alpha},
\label{pola_newbasis}
\end{align} 
with
\begin{align*}
g_{j}= \frac{\widetilde{\omega}^{2}_{j} \left(1+ \nu^{2}_{\rm pn}/\omega^{2}_{\rm pl}\right)-\omega^{2}_{0}}{\widetilde{\omega}^{2}_{j}-\widetilde{\omega}^{2}_{j'}} \sqrt{\frac{\omega^{4}_{\rm pl}}{\widetilde{\omega}^{4}_{j}}+\frac{\nu^{2}_{\rm pn}\omega^{2}_{\rm pl}}{(\widetilde{\omega}^{2}_{j}-\omega^{2}_{0})^{2}}},
\end{align*} 
$j'\neq j$, and $V=S\ell$. In the absence of phonon-photon coupling ($\nu_{\rm pn}=0$), the two modes $j=1,2$ reduce to the bare phonon and plasmon. A similar situation occurs for $\nu_{\rm pn}\neq 0$, in the case of large plasmon-phonon detunings. Both in the high $\omega_{\rm pl}\gg \omega_{0}$ and low $\omega_{\rm pl}\ll \omega_{0}$ electron density regimes, the hybrid modes reduce to the bare plasmon and phonon excitations, and the plasmon contribution prevails in the polarization field Eq.~(\ref{pola_newbasis}). As explained in the following, when the latter interacts with the photonic displacement field ${\bf D}$, this simply results in the formation of SPPs coexisting with bare phonons. The latter are either transverse phonons with frequency $\omega_{0}$ for $\omega_{\rm pl}\gg \omega_{0}$, or longitudinal phonons with frequency $\widetilde{\omega}_{0}$ in the case $\omega_{\rm pl}\ll \omega_{0}$ (Fig.~\ref{fig1}). In the following, we focus on the most interesting case occuring close to the resonance $\omega_{\rm pl}\sim \omega_{0}$, where plasmon-phonon hybridization occurs. 

\section{Coupling to photons}

Due to the breaking of translational invariance in the $z$ direction, the 3D photon wave vector can be split into an in-plane and a transverse component in each media as ${\bf Q} = q {\bf u}_{\parallelsum} + i\gamma_{n}{\bf u}_{z}$, where $n={\rm d},{\rm cr}$ refers to the dielectric medium and the crystal, respectively. For a given interface lying at the height $z_0$, the electromagnetic field associated to surface waves decays exponentially on both sides as $e^{\pm \gamma_{n}(z-z_{0})}$ with the (real) penetration depth $\gamma_{n}$. The displacement and magnetic fields can be written as superpositions of the fields generated by each interfaces $m=1,2$, namely ${\bf D} ({\bf R}) = \sum_{{\bf q},m} \sqrt{\frac{4\epsilon_{0}\hbar c}{S}} e^{i {\bf q}\cdot {\bf r}} {\bf u}_{{\bf q}m} (z) D_{{\bf q}m}$ and ${\bf H} ({\bf R}) = \sum_{{\bf q},m} w_{q} \sqrt{\frac{4\epsilon_{0}\hbar c}{S}} e^{i {\bf q}\cdot {\bf r}} {\bf v}_{{\bf q}m} (z) H_{{\bf q}m}$, with $w_{q}$ the frequency of the surface waves (still undertermined) and $q\equiv \vert {\bf q} \vert$. The mode profile functions ${\bf u}_{{\bf q}m} (z)$ and ${\bf v}_{{\bf q}m}(z)$ depending on $\gamma_{n}$ are provided in the supplemental material. Using these expressions, the photon Hamiltonian takes the form
\begin{align*}
H_{\rm pt}= \hbar c\sum_{{\bf q},m,m'} \left(\mathcal{A}^{m m'}_{q} D_{{\bf q}m} D_{{\bf -q}m'}+ \mathcal{B}^{m m'}_{q} H_{{\bf q}m} H_{{\bf -q}m'} \right), 
\end{align*} 
where the overlap matrix elements $\mathcal{A}^{m m'}_{q}$ and $\mathcal{B}^{m m'}_{q}$ depend on the parameters $\gamma_{\rm cr}$ and $\gamma_{\rm d}$. The next step corresponds to finding the electromagnetic field eigenmodes, which consist of a symmetric and an antisymmetric mode, such that the photon Hamiltonian can be put in the diagonal form:
\begin{align*}
H_{\rm pt}= \hbar c\sum_{{\bf q},\sigma=\pm} \left(\alpha_{q\sigma} D_{{\bf q}\sigma} D_{{\bf -q}\sigma}+ \beta_{q\sigma} H_{{\bf q}\sigma} H_{{\bf -q}\sigma} \right),
\end{align*}  
with $\alpha_{q\pm}=\mathcal{A}^{11}_{q} \pm \mathcal{A}^{12}_{q}$, $\beta_{q\pm}=\mathcal{B}^{11}_{q} \pm \mathcal{B}^{12}_{q}$, $D_{{\bf q}\pm}=\left(D_{{\bf q}2} \pm D_{{\bf q}1} \right)/\sqrt{2}$ and a similar expression for $H_{{\bf q}\pm}$. The new field operators $D_{{\bf q}\sigma}$ and $H_{{\bf q}\sigma}$ satisfy the commutation relations $[D_{{\bf q}\sigma},H^{\dagger}_{{\bf q'}\sigma'}]=-i C_{q\sigma} \delta_{{\bf q},{\bf q'}}\delta_{\sigma,\sigma'}$, together with the properties $D^{\dagger}_{{\bf q}\sigma}=D_{{\bf -q}\sigma}$ and $H^{\dagger}_{{\bf q}\sigma}=H_{{\bf -q}\sigma}$. The constant $C_{q\sigma}$ is determined using the Amp\`{e}re's circuital law, which provides $C_{q\sigma} = \frac{w_{q}}{2c\beta_{q\sigma}}$~\cite{todorovY}. 

The light-matter coupling Hamiltonian $H_{\rm mat-pt}$ is derived using the expression of ${\bf D} ({\bf R})$ in the new basis together with Eq.~(\ref{pola_newbasis}). While ${\bf q}$ is a good quantum number due to the in-plane translational invariance, the perpendicular wave vector of the 3D matter modes $q_{z}$ is not. In the light-matter coupling Hamiltonian $H_{\rm mat-pt}$, photon modes with a given ${\bf q}$ interact with linear superpositions of the 3D matter modes exhibiting different $q_{z}$. The latter are denoted as quasi-2D ``bright'' modes, and are defined as $\pi_{{\bf q}\sigma j}=\sum_{q_{z},\alpha} f_{\alpha\sigma} (Q) \Pi_{{\bf Q}j\alpha}$, where $f_{\alpha\sigma} (Q)$ stem from the overlap between the displacement and the polarization fields and is determined by imposing the commutation relations $[\pi_{{\bf q}\sigma j},\pi^{\dagger}_{{\bf q'}\sigma' j'}]=\delta_{{\bf q},{\bf q'}}\delta_{\sigma,\sigma'}\delta_{j,j'}$. 

Using a unitary transformation (see supplemental material), the matter Hamiltonian can be decomposed as $H_{\rm mat}=\sum_{q,\sigma,j} \hbar \widetilde{\omega}_{j} \pi^{\dagger}_{{\bf q}\sigma j} \pi_{{\bf q}\sigma j} + H_{\rm dark}$, where the second term is the contribution of the ``dark'' modes that are orthogonal to the bright ones, and which \textit{do not interact with photons.} Without this contribution, the polariton Hamiltonian reads $H_{\rm pol}=\sum_{q,\sigma} \mathcal{H}_{q\sigma}$ with
\begin{align*}
\mathcal{H}_{q\sigma} = \hbar c \left(\alpha_{q\sigma} D_{{\bf q}\sigma} D_{{\bf -q}\sigma}+ \beta_{q\sigma} H_{{\bf q}\sigma} H_{{\bf -q}\sigma} \right) + \sum_{j} \hbar \widetilde{\omega}_{j} \pi^{\dagger}_{{\bf q}\sigma j} \pi_{{\bf q}\sigma j} - \sum_{j} \hbar \Omega_{q \sigma j} \left(\pi_{{\bf -q}\sigma j} + \pi^{\dagger}_{{\bf q}\sigma j}\right) D_{{\bf q}\sigma},
\end{align*}
and the vacuum Rabi frequency 
\begin{align*}
\Omega_{q \sigma j} = g_{j} \sqrt{\frac{c \omega^{2}_{\rm pl}}{\widetilde{\omega}_{j}}}\sqrt{\frac{q^{2}+\gamma^{2}_{\rm cr}}{\gamma_{\rm cr}} \left(1-e^{-2 \gamma_{\rm cr}\ell}\right) + 2 \ell \sigma e^{-\gamma_{\rm cr}\ell}\left(q^{2}-\gamma^{2}_{\rm cr} \right)}.
\end{align*} 

\section{Surface plasmon-phonon polaritons}

The Hamiltonian $H_{\rm pol}$ exhibits three eigenvalues in each subspace $({\bf q},\sigma)$, and only the lowest two (refered to as lower and upper polaritons) correspond to surface modes (below the light cone). At this point, we have built an Hamiltonian theory providing a relation between the field penetration depths $\gamma_{n}$ and the surface wave frequencies $w_{q\sigma\zeta}$, where $\zeta={\rm LP},{\rm UP}$ refers to the lower (LP) and upper (UP) polaritons. This eigenvalue equation can be combined with the Helmholtz equation $\epsilon_{n} (\omega) \omega^{2}/c^{2}=q^{2}-\gamma^{2}_{n}$ in order to determine these parameters~\cite{todorovY}. We use a self-consistent algorithm which starts with a given frequency $w_{q\sigma{\rm LP}}$, then determine $\gamma_{n}$ from the Helmholtz equation with $\epsilon_{\rm d}=1$ and $\epsilon_{\rm cr} (w_{q\sigma{\rm LP}})$ given by Eq.~(\ref{epsi_cr}), and use these $\gamma_{n}$ to compute the parameters entering the Hamiltonian $H_{\rm pol}$. The latter is diagonalized numerically (see supplemental material), which allows to determine the new $w_{q\sigma{\rm LP}}$. The algorithm is applied independently for the symmetric ($\sigma=+$) and antisymmetric ($\sigma=-$) modes until convergence, which is ensured by the discontinuity of both light and matter fields at each interface~\cite{todorovY}. 

\clearpage

\begin{figure}[b!]
\centerline{\includegraphics[width=0.95\columnwidth]{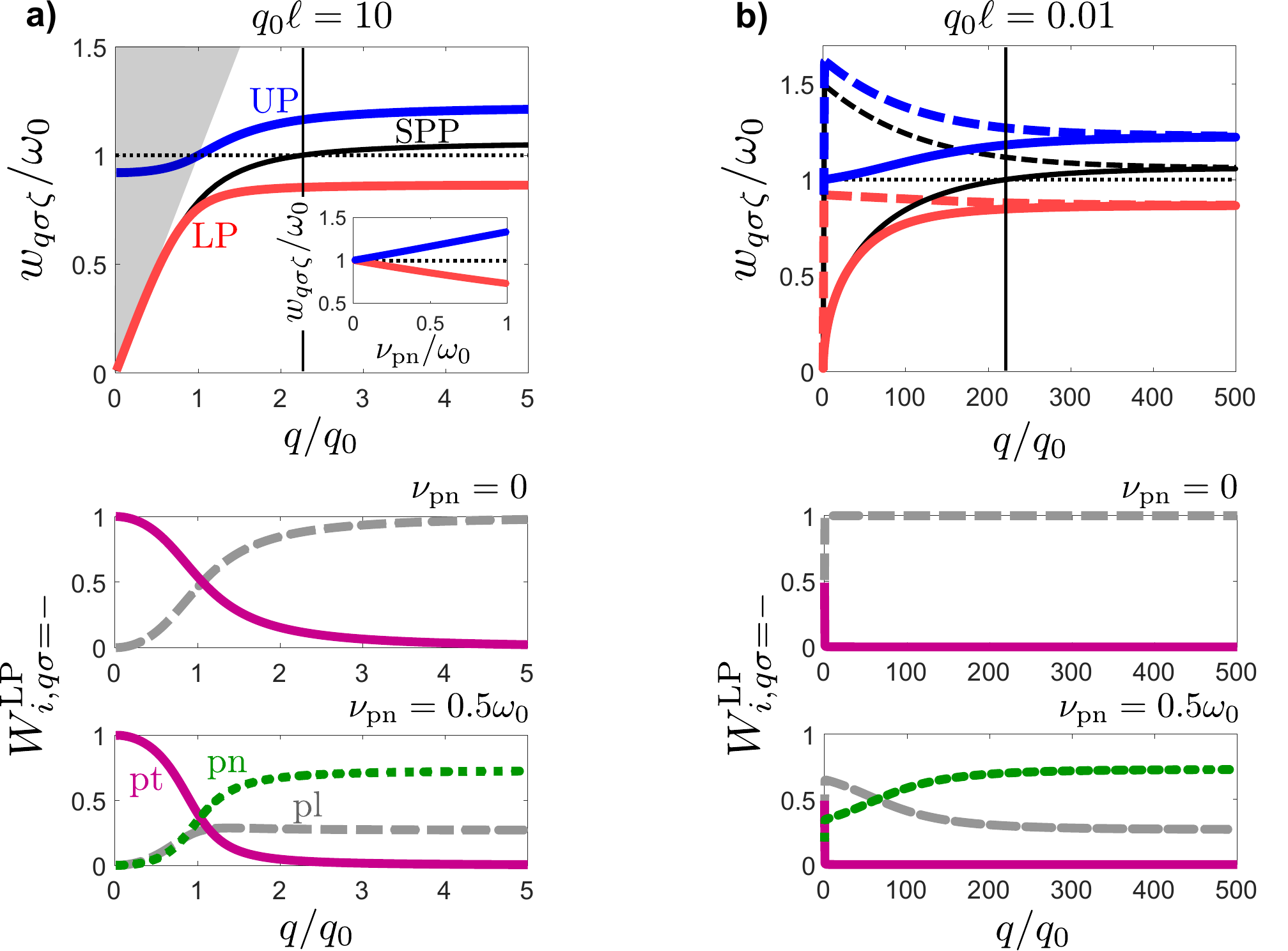}}
\caption{Normalized frequency dispersion $w_{q\sigma\zeta}/\omega_{0}$ and mode admixtures of the surface polaritons as a function of $q/q_{0}$ for $\omega_{\rm pl}=1.5\omega_{0}$. \textbf{a)} $q_{0} \ell = 10$. \textbf{b)} $q_{0} \ell = 0.01$. Top panel: The SPPs obtained for $\nu_{\rm pn}=0$ are depicted as black lines, while the surface lower and upper polaritons obtained for $\nu_{\rm pn}=0.5\omega_{0}$ are represented as thick red and blue lines, respectively. Resonance between SPPs and phonons (horizontal dotted lines) is indicated by thin vertical lines. The antisymmetric ($\sigma=-$) and symmetric ($\sigma=+$) modes correspond to the solid and dashed lines, and the light cone with boundary $\omega= q c$ is represented a grey shaded region. Inset: Frequency $w_{q\sigma\zeta}/\omega_{0}$ of the lower (red line) and upper (blue line) surface polaritons versus $\nu_{\rm pn}/\omega_{0}$ for $q_{0} \ell = 10$ and $\omega_{\rm pl}=1.5\omega_{0}$. The phonon frequency is depicted as an horizontal dotted line. Bottom panels: Plasmon (grey dashed lines), photon (magenta solid lines), and phonon (green dotted lines) admixtures of the antisymmetric lower polariton $W^{\rm LP}_{i,q\sigma=-}$ ($i={\rm pl},{\rm pt},{\rm pn}$) as a function of $q/q_{0}$ for $\omega_{\rm pl}=1.5\omega_{0}$. The top and bottom sub-panels correspond to $\nu_{\rm pn}=0$ and $\nu_{\rm pn}=0.5\omega_{0}$, respectively.} 
\label{fig4}
\end{figure}

We now use this method to study the surface polaritons in our system. As an example, we consider the case $\omega_{\rm pl}=1.5\omega_{0}$ with different crystal thickness $q_{0}\ell=10$ [Fig.~\ref{fig4} \textbf{a)}] and $q_{0}\ell=0.01$ [Fig.~\ref{fig4} \textbf{b)}], and compute the surface polariton frequencies $w_{q\sigma\zeta}$ (top panels), as well as their phonon ($i={\rm pn}$), plasmon ($i={\rm pl}$), and photon ($i={\rm pt}$) admixtures $W^{\rm LP}_{i,q\sigma}$ for $\nu_{\rm pn}=0$ and $\nu_{\rm pn}=0.5\omega_{0}$ (bottom panels). Precise definitions of these quantities are provided in the supplemental material, and $\ell$ and $q$ are both normalized to $q_{0}=\omega_{0}/c$. Considering typical mid-infrared phonons with $\hbar \omega_{0}\sim 0.2 {\rm eV}$, the two dimensionless parameters $q_{0}\ell=10$ and $q_{0}\ell=0.01$ correspond to $\ell = 10 {\rm \mu m}$ (first case) and $\ell \sim 10 {\rm nm}$ (second case), respectively. 

In the first case [Fig.~\ref{fig4} \textbf{a)}], the two surface modes at each interface have negligible overlap and the modes $\sigma_{\pm}$ therefore coincide. For $\nu_{\rm pn}=0$, the plasmon-photon coupling is responsible for the appearance of a SPP mode (black line) with frequency $w^{0}_{q}$, which enters in resonance with the phonon mode at $q_{\rm r}\approx 2.2 q_{0}$. While for $q \ll q_{0}$ this SPP is mainly composed of light ($w^{0}_{q} \sim q c$), it features an hybrid plasmon-photon character for $q \sim q_{0}$, and becomes mostly plasmon-like as $w^{0}_{q}$ approaches the surface plasmon frequency $\omega_{\rm pl}/\sqrt{2}$ for $q \gg q_0$. For $\nu_{\rm pn}\neq 0$, a splitting between the two polaritons branches (thick red and blue lines) is clearly visible, and the latter consist of a mix between phonons, plasmons, and photons in the vicinity of $q=q_{0}$. In the regime $q > q_{0}$, since the mostly phonon-like LP exhibits a $\sim 25\%$ plasmon admixture, this polariton mode can be seen as a ``dressed'' phonon with frequency red-shifted from $\omega_{0}$. Similarly, the ``dressed'' plasmon mode (UP) is blue-shifted from the surface mode frequency $\omega_{\rm pl}/\sqrt{2}$ due its $\sim 25\%$ phonon weight.

The frequencies of the LP (red line) and UP (blue line) at resonance are represented as a function of $\nu_{\rm pn}/\omega_{0}$ on the inset. Here, the resonance is defined by the condition $w^{0}_{q} =\omega_{0}$, which provides $q_{\rm r}\approx 2.2 q_{0}$. We observe that the polariton splitting $w_{q\sigma{\rm UP}}-w_{q\sigma{\rm LP}}$ is symmetric with respect to $w_{q\sigma\zeta}=\omega_{0}$ (black dotted line) for $\nu_{\rm pn}\ll \omega_{0}$, and becomes slightly asymmetric in the ultrastrong coupling regime\cite{ciuti1}, when $\nu_{\rm pn}$ is a non-negligible fraction of $\omega_{0}$. Furthermore, the splitting is found to decrease rapidly as the electronic plasma frequency $\omega_{\rm pl}$ is increased.

In the second case $q_{0}\ell=0.01$ [Fig.~\ref{fig4} \textbf{b)}], the crystal is thinner than the penetration depth $\gamma_{\rm cr}$ of the surface waves at the two interfaces, and the latter overlap. This results in two sets of modes $\sigma=\pm$ with different frequencies. For $\nu_{\rm pn}=0$, the symmetric and antisymmetric SPPs with frequency $w^{0}_{q\sigma=\pm}$ are represented as black dashed and solid lines, respectively. The resonance between the symmetric SPP and the phonon mode occurs for $q\sim q_0$, in the regime where the symmetric SPP is mostly photon-like with $w^{0}_{q\sigma=+}\sim q c$. Interestingly, the resonance between the antisymmetric SPP and the phonon mode is now shifted to a large wave vector $q_{\rm r} \approx 220 q_{0}$, where the antisymmetric SPP exhibits a pure plasmonic character. For $q/q_{0} \to \infty$, the two SPPs $\sigma=\pm$ converge to the surface plasmon frequency $\omega_{\rm pl}/\sqrt{2}$. For $\nu_{\rm pn}\neq 0$, the antisymmetric LP (thick red solid line) and UP (thick blue solid line) are splitted in the vicinity of the resonance $q_{\rm r} \approx 220 q_{0}$, while the symmetric polaritons (thick colored dashed lines) do not feature any anticrossing behavior for $q\sim q_0$. Similarly as in \textbf{a)}, the LPs $\sigma=\pm$ can be seen as dressed phonon modes with frequencies red-shifted from $\omega_{0}$ for large $q$ due to their plasmon admixtures. Here, however, these dressed phonons can propagate with very large wave vectors comparable to the electronic Fermi wave vector.   

\section{Electron-phonon scattering}

\begin{figure}[ht].
\centerline{\includegraphics[width=0.95\columnwidth]{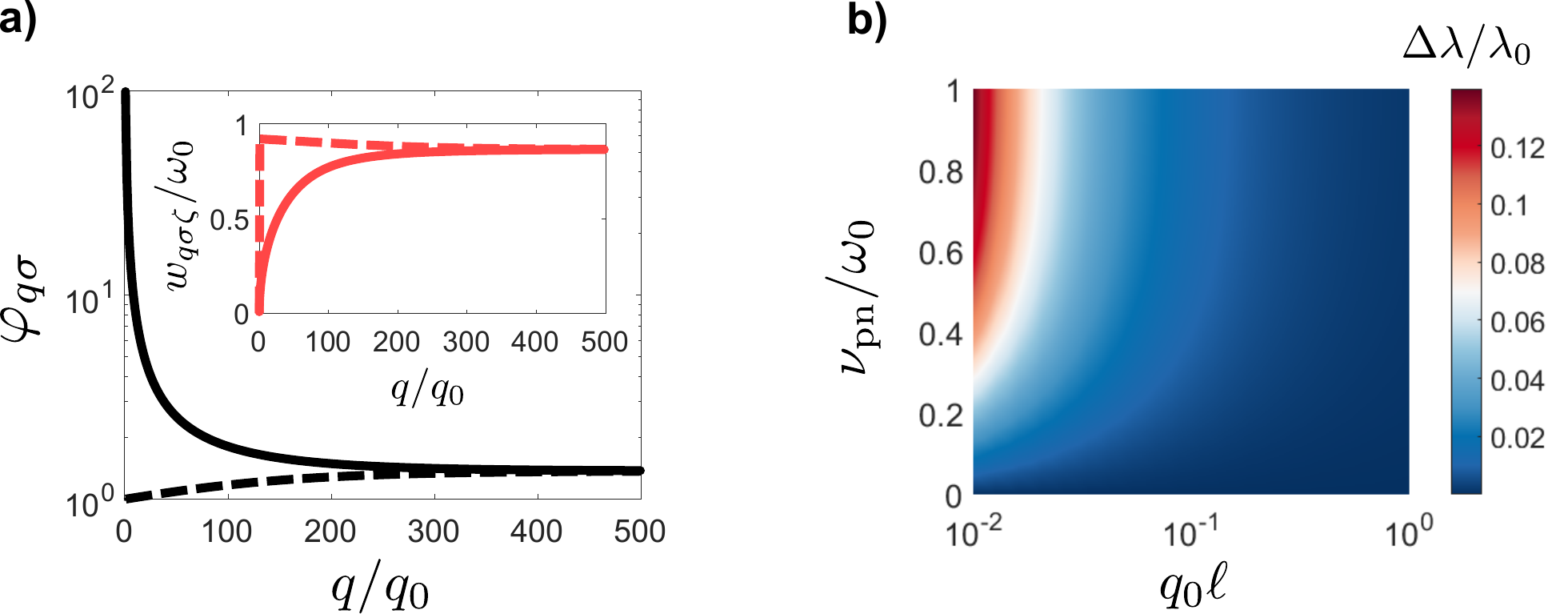}}
\caption{\textbf{a)} Function $\varphi_{q\sigma}$ versus $q/q_0$ for $\nu_{\rm pn}/\omega_{0}=0.5$, $q_{0} \ell = 0.01$, and $\omega_{\rm pl}=1.5\omega_{0}$. Inset: Normalized frequency dispersion of the LPs $w_{q\sigma{\rm LP}}/\omega_{0}$ corresponding to the red solid and dashed lines on Fig.~\ref{fig4} \textbf{b)}. The contributions of symmetric ($\sigma=+$) and antisymmetric ($\sigma=-$) modes are depicted as dashed and solid lines, respectively. \textbf{b)} Relative enhancement of the electron-phonon coupling parameter $\frac{\Delta\lambda}{\lambda_{0}}$ given by Eq.~(\ref{self_renorm}) versus $q_{0}\ell$ and $\nu_{\rm pn}/\omega_{0}$, for $\omega_{\rm pl}=1.5\omega_{0}$ and $K_{\rm F}=10^{3} q_{0}$.}  
\label{fig5}
\end{figure}

We now show that the ability to tune the energy and wave vector of the dressed phonons can affect the electron-phonon scattering in the crystal. The coupling Hamiltonian between individual electrons (electronic dark modes) and phonons reads
\begin{align}
H_{\rm el-pn}= \sum_{\bf K} \hbar \xi_{K} c^{\dagger}_{\bf K} c_{\bf K} + \sum_{{\bf K},{\bf Q},\alpha} \hbar \mathcal{M} c^{\dagger}_{\bf K} c_{{\bf K}-{\bf Q}} \left(B_{{\bf Q}\alpha}+ B^{\dagger}_{{\bf -Q}\alpha}\right). 
\label{hami_ph_el} 
\end{align}
The fermionic operators $c_{\bf K}$ and $c^{\dagger}_{\bf K}$ annihilate and create an electron with wave vector ${\bf K}$ and energy $\hbar \xi_{K}$ ($K\equiv \vert {\bf K}\vert$) relative to the Fermi energy. For simplicity, we assume a 3D spherical Fermi surface providing $\xi_{K}=\hbar (K^{2}-K^{2}_{\rm F})/(2 m)$, with $m$ the electron effective mass and $K_{\rm F}$ the Fermi wave vector, and that the coupling constant $\mathcal{M}$ does not depend on ${\bf Q}$ as shown theoretically for intramolecular phonons in certain crystals~\cite{PhysRevB.58.8236}. Electron-phonon interactions are usually characterised by the dimensionless coupling parameter $\lambda$, which quantifies the electron mass renormalization due to the coupling to phonons~\cite{mahan}. At zero temperature, $\lambda$ is defined as
\begin{align*}
\lambda= \frac{1}{N_{\rm 3D}}\sum_{\bf K} \delta (\xi_{K}) \Re \left(-\partial_{\omega} \overline{\Sigma}_{\bf K} (\omega) \vert_{\omega=0} \right),
\end{align*}
where $\Re$ stands for real part, $N_{\rm 3D}=\sum_{\bf K} \delta (\xi_{K})=\frac{V m K_{\rm F}}{2 \pi^2 \hbar}$ is the 3D electron density of states at the Fermi level, and $\partial_{\omega} \overline{\Sigma}_{\bf K} (\omega) \vert_{\omega=0}$ denotes the frequency derivative of the retarded electron self-energy $\overline{\Sigma}_{\bf K} (\omega)$ evaluated at $\omega=0$. An equation of motion analysis~\cite{PhysRev.131.993} of the electron Green's function (GF) $\mathcal{G}_{\bf K} (\tau)=-i \langle c_{\bf K} (\tau) c^{\dagger}_{\bf K} (0) \rangle$ allows to write the electron self-energy as
\begin{align}
\Sigma_{\bf K}(\omega)= \sum_{{\bf Q},\alpha} i \vert \mathcal{M} \vert^{2} \int \!\! \frac{d\omega'}{2\pi} \mathcal{G}_{{\bf K}-{\bf Q}} (\omega+\omega') \mathfrak{B}_{Q\alpha} (\omega'), 
\label{self_eq0}
\end{align}
where $\mathfrak{B}_{Q\alpha} (\omega)=-i\int \! d\tau e^{i\omega \tau} \langle \mathcal{B}_{{\bf Q}\alpha} (\tau)\mathcal{B}_{{\bf -Q}\alpha} (0) \rangle$ is the phonon GF written in the frequency domain, and $\mathcal{B}_{{\bf Q}\alpha}=B_{{\bf Q}\alpha}+B^{\dagger}_{{\bf -Q}\alpha}$. In the absence of phonon-photon coupling ($\nu_{\rm pn}=0$), there is no hybridization between phonons and plasmons, nor is there coupling to photons. Phonons therefore enter the Hamiltonian $H_{\rm pol}$ only in the free contribution $H_{\rm pn}=\sum_{{\bf Q},\alpha}\hbar\omega_{0}B^{\dagger}_{{\bf Q}\alpha}B_{{\bf Q}\alpha}$. In this case, the equation of motion analysis simply provides $\mathfrak{B}_{Q\alpha}(\omega)=2\omega_{0}/(\omega^{2}-\omega^{2}_{0})$. Using this expression together with the non-interacting electron GF $\mathcal{G}^{0}_{\bf K} (\omega)=1/(\omega-\xi_{K})$ in Eq.~(\ref{self_eq0}), one can compute the electron-phonon coupling parameter for $\nu_{\rm pn}=0$ as (see supplemental material)
\begin{align*}
\lambda_{0}= \frac{2 \vert \mathcal{M}\vert^{2}}{N_{\rm 3D}\omega_{0}} \sum_{\bf K} \delta (\xi_{K}) \sum_{\bf Q} \delta (\xi_{{\bf K}-{\bf Q}}) =\frac{2 \vert \mathcal{M} \vert^{2} N_{\rm 3D}}{\omega_{0}}.
\end{align*}
In the presence of phonon-photon coupling ($\nu_{\rm pn}\neq 0$), the phonon dynamics is governed by the Hamiltonian $H_{\rm pol}$, which includes the coupling of phonons to plasmons and photons. It is therefore convenient to express the 3D phonon operators $B_{{\bf Q}\alpha}$ and $B^{\dagger}_{{\bf Q}\alpha}$ in terms of the 3D hybrid modes $\Pi_{{\bf Q}\alpha j}$ and $\Pi^{\dagger}_{{\bf Q}\alpha j}$, and then project the latter onto the quasi-2D bright and dark modes such that the electron-phonon Hamiltonian Eq.~(\ref{hami_ph_el}) takes the form $H_{\rm el-pn}=\sum_{\bf K} \hbar \xi_{K} c^{\dagger}_{\bf K} c_{\bf K}+ H^{\rm (B)}_{{\rm el}-{\rm pn}}+H^{\rm (D)}_{{\rm el}-{\rm pn}}$. The contribution of the bright modes reads
\begin{align*}
H^{\rm (B)}_{{\rm el}-{\rm pn}}= \sum_{{\bf K},{\bf Q}} \sum_{\alpha,\sigma,j} \hbar \mathcal{M} \eta_{j} f^{*}_{\alpha\sigma} (Q) c^{\dagger}_{\bf K} c_{{\bf K}-{\bf Q}} \left(\pi_{{\bf q}\sigma j}+ \pi^{\dagger}_{{\bf -q}\sigma j} \right),
\end{align*}
while the contribution $H^{\rm (D)}_{{\rm el}-{\rm pn}}$ of the dark modes that do not interact with photons is given in the supplemental material. Here, $\eta_{j}= \chi_{j} \sqrt{\frac{\omega_{0}/\widetilde{\omega}_{j}}{(\omega_{\rm pl}/\widetilde{\omega}_{j})^{4} + \chi_{j}^{2}}}$ with $\chi_{j}=\frac{\nu_{\rm pn}\omega_{\rm pl}}{\widetilde{\omega}^{2}_{j}-\omega^{2}_{0}}$ is associated to the hybrid modes weights of the phonons. The electron self-energy due to the interaction with bright modes reads 
\begin{align}
\Sigma^{\rm (B)}_{\bf K} (\omega)= \sum_{{\bf Q},\alpha,\sigma,j} i \eta^{2}_{j} \vert \mathcal{M} f_{\alpha\sigma} (Q) \vert^{2} \int \!\! \frac{d\omega'}{2\pi} \mathcal{G}_{{\bf K}-{\bf Q}} (\omega+\omega') \mathfrak{P}_{q\sigma j} (\omega'),
\label{self_eq236}
\end{align}
where $\mathfrak{P}_{q\sigma j} (\omega)=-i\int \! d\tau e^{i\omega \tau} \langle \Upsilon_{{\bf q}\sigma j} (\tau) \Upsilon_{{\bf -q}\sigma j} (0) \rangle$ and $\Upsilon_{{\bf q}\sigma j}=\pi_{{\bf q}\sigma j}+\pi^{\dagger}_{{\bf -q}\sigma j}$. As detailed in the supplemental material, we now use the equation of motion theory to calculate the GF $\mathfrak{P}_{q\sigma j} (\omega)$. For $\nu_{\rm pn}=0$, one simply obtains $\mathfrak{P}_{q\sigma j} (\omega)=2\widetilde{\omega}_{j}/(\omega^{2}-\widetilde{\omega}^{2}_{j})$, which provides $\mathfrak{P}_{q\sigma 1} (\omega)=\mathfrak{B}_{Q\alpha}(\omega)$ since $\widetilde{\omega}_{1}=\omega_{0}$ (and $\widetilde{\omega}_{2}=\omega_{\rm pl}$). In this case, $\eta_{1}=1$, $\eta_{2}=0$, and only the pure phonon term $j=1$ therefore contributes to $\Sigma^{\rm (B)}_{\bf K}$. We then calculate $\mathfrak{P}_{q\sigma j} (\omega)$ and the self-energy Eq.~(\ref{self_eq236}) for $\nu_{\rm pn}\neq 0$, and decompose the electron-phonon coupling parameter into its bright and dark mode contributions: $\lambda=\lambda^{\rm (B)}+\lambda^{\rm (D)}$ for $\nu_{\rm pn}\neq 0$, and $\lambda_{0}=\lambda^{\rm (B)}_{0}+\lambda^{\rm (D)}_{0}$ for $\nu_{\rm pn}= 0$. The contributions $\lambda^{\rm (B)}$ and $\lambda^{\rm (B)}_{0}$ are different because bright modes interact with photons for $\nu_{\rm pn}\neq 0$, while they do not for $\nu_{\rm pn}= 0$. Similarly, $\lambda^{\rm (D)}$ and $\lambda^{\rm (D)}_{0}$ are \textit{a priori} different since dark modes are hybrid plasmon-phonon modes for $\nu_{\rm pn}\neq 0$, while they reduce to bare phonons for $\nu_{\rm pn}= 0$. However, we show in the supplemental material that $\lambda^{\rm (D)}=\lambda^{\rm (D)}_{0}$, which means that, at the level of the RPA, the hybridization between plasmons and phonons can not solely lead to a modification of the electron-phonon scattering. Finally, the relative enhancement of the electron-phonon coupling parameter reads (see supplemental material) 
\begin{align}
\frac{\Delta\lambda}{\lambda_{0}}\equiv \frac{\lambda-\lambda_{0}}{\lambda_{0}}= \frac{1}{N^{2}_{\rm 3D}} \sum_{\bf K} \delta (\xi_{K}) \sum_{{\bf Q},\alpha,\sigma} \left(\varphi_{q\sigma} - 1 \right) \vert f_{\alpha\sigma} (Q) \vert^{2} \delta (\xi_{{\bf K}-{\bf Q}}),
\label{self_renorm}
\end{align}
where the function $\varphi_{q\sigma}$ is given in the supplemental material. The latter describes the renormalization of the phonon energy due to the coupling to plasmons and photons, and depends on the polariton frequencies $w_{q\sigma\zeta}$. It is represented on Fig.~\ref{fig5} \textbf{a)} together with the frequency dispersion of the LPs $w_{q\sigma{\rm LP}}/\omega_{0}$ corresponding to the red solid and dashed lines on Fig.~\ref{fig4} \textbf{b)}. First, we find that the contributions of the LPs to $\varphi_{q\sigma}$ largely dominate over the contributions of the other polaritons. Secondly, we observe that $\varphi_{q\sigma}$ is anticorrelated with the dispersion of the LPs, whenever the latter exhibit a finite phonon weight. For instance, $\varphi_{q\sigma=-}$ reaches large values in the region $q \lesssim q_{0}$ where the frequency of the antisymmetric LP ($\sim 35\%$ phonon for $\nu_{\rm pn} =0.5\omega_{0}$) is far away from the bare phonon frequency. In contrast, while $w_{q\sigma=+{\rm LP}} \ll \omega_{0}$ in this region, $\varphi_{q\sigma=+} \approx 1$ since the symmetric LP is fully photon-like for $q \lesssim q_{0}$. Note that for $\nu_{\rm pn}/\omega_{0} \to 0$, the LPs for $q > q_{\rm r}$ are fully phonon-like with $w_{q\sigma{\rm LP}} \to \omega_{0}$, and one finds that $\varphi_{q\sigma} \to 1$ does not contribute to $\frac{\Delta\lambda}{\lambda_{0}}$ in this region [see Eq.~(\ref{self_renorm})]. Similarly, $\varphi_{q\sigma}$ does not contribute to $\frac{\Delta\lambda}{\lambda_{0}}$ for $q< q_{\rm r}$ because of the vanishing phonon weight of the LPs for $\nu_{\rm pn}/\omega_{0} \to 0$.        

\clearpage

\noindent The relative enhancement of the electron-phonon coupling $\frac{\Delta\lambda}{\lambda_{0}}$ is represented on Fig.~\ref{fig5} \textbf{b)} as a function of $q_{0}\ell$ and $\nu_{\rm pn}/\omega_{0}$. We find that it can reach large values $\gtrsim 10\%$ for $\nu_{\rm pn}\gtrsim 0.4\omega_{0}$ and $\ell$ sufficiently small ($q_{0}\ell \sim 0.01$). The fact that $\frac{\Delta\lambda}{\lambda_{0}}$ decreases when increasing the crystal thickness $\ell$ can be understood by realizing that the function $\vert f_{\alpha\sigma} (Q) \vert^{2}$ entering Eq.~(\ref{self_renorm}) scales as $\sim 1/(\gamma_{\rm cr} \ell)$, as shown in the supplemental material. This means that the dressing of 3D phonons by evanescent waves decaying exponentially at each interface is large for thin enough crystals, when surface effects become important. Nevertheless, when the crystal thickness becomes too small, namely when $K_{\rm F}\ell \lesssim 1$, the description of the matter excitations by 3D fields breaks down and quantum confinement effects have to be taken into account in the model. The parameters of Fig.~\ref{fig5} \textbf{b)} are $\ell = 10 {\rm nm}$ for $\hbar \omega_{0}=0.2 {\rm eV}$, and $K_{\rm F} = 1 {\rm nm}^{-1}$. We find that the enhancement of the electron-phonon coupling strongly increases with the ratio $q_{0}/K_{\rm F}$, i.e. when the mismatch between the typical photonic and electronic wave vectors is reduced. This can occur in materials with larger phonon frequency and/or lower electron density. In the latter case, however, decreasing $K_{\rm F}$ requires to consider larger $\ell$ for the 3D description of the matter fields to be valid, which in turn leads to a reduction of $\frac{\Delta\lambda}{\lambda_{0}}$.  

For instance, bilayer graphene with $\ell \sim 0.7 {\rm nm}$~\cite{Razado} exhibits an infrared-active phonon mode at $\hbar \omega_{0}= 0.2 {\rm eV}$~\cite{PhysRevLett.103.116804_2009}, and has been recently shown to feature superconductivity at low carrier densities $\approx 2\times 10^{12} {\rm cm}^{-2}$ when the two sheets of graphene are twisted relative to each other by a small angle~\cite{yuan}. In this regime, we find that $\omega_{\rm pl}\approx \omega_{0}$, $q_{0} \ell \approx 7 \times 10^{-4}$, $K_{\rm F}\approx 1.6 \times 10^{3} q_{0}$, and the large phonon-photon coupling strengths $\nu_{\rm pn}\approx 0.25\omega_{0}$~\cite{PhysRevLett.103.116804_2009} allow to reach the ultrastrong coupling regime. Using these parameters, we find that the contribution of surface plasmon-phonon polaritons to the electron-phonon scattering is expected to reach a very large value $\frac{\Delta\lambda}{\lambda_{0}} \sim 2.7$. However, the small thickness of bilayer graphene such that $K_{\rm F}\ell \sim 1$ in this weakly-doped regime implies that quantum confinement effects become important. Furthemore, the failure of the 3D model assuming $\omega_{\rm pl} \to \textrm{const}$ as $q\to 0$ can be seen in the dispersion of the optical plasmon mode in bilayer graphene~\cite{PhysRevB.82.195406}, which is known to be $\omega_{\rm pl}\sim \sqrt{q}$ as $q\to 0$. 

\section{Conclusion}

In summary, we have carried out a non-perturbative quantum theory of intrinsic, lossless plasmon-phonon polaritons valid for arbitrary large coupling strengths, and discussed the different regimes of interest obtained by tuning the crystal thickness and the phonon-photon coupling strength. This theory allows to characterize the hybridization between plasmons and phonons in the crystal, which can not lead solely to a modification of the electron-phonon scattering at the level of the RPA. Considering the coupling of these hybrid plasmon-phonon modes to photons, we have found that in the regime where both the electronic and ionic plasma frequencies become comparable to the phonon frequency, a dressed low-energy phonon mode with finite plasmon weight arises. In the case of quasi-2D crystals, the in-plane wave vector of this dressed phonon can reach very large values comparable to the Fermi wave vector, which is shown to lead to an enhancement of the electron-phonon scattering. It is an interesting prospect to investigate whether this effect could modify superconductivity in certain molecular crystals. While our model specifically adresses the case of metallic crystals with large ionic charges supporting far/mid-infrared phonons~\cite{PhysRevB.16.3283_1977,doi:10.1063/1.437662_1979,PhysRevB.45.10173_1992,PhysRevB.46.1937_1992,PhysRevLett.103.116804_2009}, it can be directly generalized to describe other type of excitations such as excitons, and extended to other geometries. Further useful extensions include the presence of quantum confinement and dissipation. In addition to the intrinsic phonon linewidth, the main source of dissipation is the damping of plasmons with large wave vectors due to the electron-hole continuum (electronic dark modes)~\cite{mahan}. However, since the dressed phonons feature only a $\sim 25\%$ plasmon admixture at large wave vectors, this effect is not expected to dramatically affect our results.  

\begin{acknowledgement}
We gratefully acknowledge discussions with M. Antezza, T. Chervy, V. Galitski, Y. Laplace, Stefan Sch\"utz, A. Thomas, and Y. Todorov. This work is partially supported by the ANR - ``ERA-NET QuantERA'' - Projet ``RouTe'' and Labex NIE.
\end{acknowledgement}

\newpage
\thispagestyle{empty}
\mbox{}
\pagebreak

\section{Supplemental material}

In this supplemental material, we provide details of the calculations sketched in the main text. The first section is dedicated to the derivation and diagonalization of the matter Hamiltonian leading to the hybrid plasmon-phonon modes. In the second section, we derive the photon and light-matter coupling Hamiltonians, and explain the self-consistent method used to find the surface polariton modes of the system. Further details concerning the theoretical foundations of the model can be found in Ref.~[\onlinecite{todorovY}]. The plasmon, phonon, and photon admixtures of the polaritons are provided at the end of the section. In the third section, we derive the coupling Hamiltonian between electronic dark modes and intramolecular optical phonons, and show in detail how the surface polariton modes can lead to an enhancement of the electron-phonon scattering in the crystal. In the last section, we study an elementary plasmonic structure composed of a single metal-dielectric interface, where phonons in the dielectric region strongly interact with SPPs. We use an effective model similar to that of Ref.~[\onlinecite{tudela_2013}], which is based on the SPP quantization scheme detailed in Refs.~[\onlinecite{PhysRevB.4.4129}-\nocite{kimMS}\onlinecite{PhysRevB.82.035411}], and show that unphysical behaviors appear in the ultrastrong coupling (USC) regime.

\subsection{Diagonalization of the matter Hamiltonian} 
\label{effe_diel}

We first explain the diagonalization procedure of $H_{\rm mat}$ in detail. The latter involves both plasmonic and phononic degrees of freedom through the total polarization field ${\bf P}={\bf P}_{\rm pl}+{\bf P}_{\rm pn}$, where
\begin{align}
{\bf P}_{\rm pl} ({\bf R})=\frac{1}{V} \sum_{{\bf K},{\bf Q}} {\bf d}_{{\bf K}{\bf Q}} \left(P_{{\bf -K}{\bf -Q}} + P^{\dagger}_{{\bf K}{\bf Q}}\right) e^{-i {\bf Q}\cdot {\bf R}}
\label{elec_pol32}
\end{align}
denotes the plasmon polarization field and ${\bf P}_{\rm pn}$ the phonon polarization. In the case of a 3D free electron gas, the electron-hole excitation operator $P^{\dagger}_{{\bf K}{\bf Q}}=c^{\dagger}_{{\bf K}+{\bf Q}} c_{\bf K}$ is written in terms of the fermionic operators $c_{\bf K}$ and $c^{\dagger}_{\bf K}$, which annihilates and creates an electron with 3D wave vector ${\bf K}$, respectively. For long-wavelength excitations with $Q \ll K_{\rm F}$, the Hilbert space is restricted to the RPA subspace spanned by the states $\{\ket{F},P^{\dagger}_{{\bf K}{\bf Q}}\ket{F}\}$, where $\ket{F}=\prod_{K< K_{\rm F}} c^{\dagger}_{\bf K} \ket{0}$ is the electron ground state (Fermi sea of the free electron gas), $K=|{\bf K}|$ the wave vector modulus, $\ket{0}$ the vacuum state, and $K_{\rm F}$ the Fermi wave vector. In this case, it can be shown~\cite{todorovY} that $P^{\dagger}_{{\bf K}{\bf Q}}$ and its hermitian conjugate are bosonic operators satisfying the commutation relation $[P_{{\bf K}{\bf Q}},P^{\dagger}_{{\bf K'}{\bf Q'}}]=\delta_{{\bf K},{\bf K'}}\delta_{{\bf Q},{\bf Q'}}$. The electronic dipole moment reads 
\begin{align*}
{\bf d}_{{\bf K}{\bf Q}}=-i e n_{\bf K} (1- n_{{\bf K}+{\bf Q}})\frac{2{\bf K}+ {\bf Q}}{(2{\bf K}+ {\bf Q})\cdot {\bf Q}},
\end{align*}
where $n_{\bf K}$ denotes the Fermi occupation number at zero temperature, namely $n_{\bf K}=1$ for $K < K_{\rm F}$ and $n_{\bf K}=0$ for $K > K_{\rm F}$. The electron effective charge and mass are denoted as $e$ and $m$, respectively. Denoting the lattice site positions as ${\bf R}_{i}$, where $i \in [1,n]$ with $n$ the number of unit cell in the crystal, the phonon polarization field can be written as
\begin{align}
{\bf P}_{\rm pn} ({\bf R})=\sqrt{\frac{\mathcal{Z}^{2} N \hbar}{2 M\omega_{0}}} \sum_{{\alpha},i} \delta \left({\bf R}- {\bf R}_{i}\right) \left(B_{i\alpha} + B^{\dagger}_{i\alpha}\right) {\bf u}_{\alpha}.
\label{pho_pol}
\end{align}
Here, ${\bf u}_{\alpha}$ with $\alpha=z,\parallelsum$ denotes the polarization (unit) vector of each phonon mode with the same frequency $\omega_{0}$, $N$ the number of vibrating ions with effective mass $M$ and charge $\mathcal{Z}$ per unit cell, and $B_{i\alpha}$ and $B^{\dagger}_{i\alpha}$ respectively annihilates and creates a phonon polarized in the direction $\alpha$ at the lattice position $i$. These operators satisfy the bosonic commutation relation $[B_{i\alpha},B^{\dagger}_{i'\alpha'}]=\delta_{\alpha,\alpha'}\delta_{i,i'}$. Writing the phonon operators in the Fourier basis: $B_{i\alpha} = \frac{1}{\sqrt{n}} \sum_{\bf Q} B_{{\bf Q}\alpha} e^{-i {\bf Q}\cdot{\bf R}_{i}}$, the phonon polarization Eq.~(\ref{pho_pol}) takes the form
\begin{align}
{\bf P}_{\rm pn} ({\bf R})=\sqrt{\frac{\mathcal{Z}^{2} N \hbar}{2 M\omega_{0} V a^{3}}} \sum_{{\bf Q},\alpha} \left(B_{{\bf -Q}\alpha} + B^{\dagger}_{{\bf Q}\alpha}\right) {\bf u}_{\alpha} e^{-i {\bf Q}\cdot {\bf R}},
\label{pho_pol22}
\end{align}
with $a^{3}$ the volume of a unit cell. The matter Hamiltonian can be decomposed as $H_{\rm mat}=H_{\rm pl}+H_{\rm pn}+H_{\rm P^{2}}$. The contribution $H_{\rm pl}=\sum_{{\bf K},{\bf Q}} \hbar \Delta \xi_{{\bf K},{\bf Q}} P^{\dagger}_{{\bf K}{\bf Q}} P_{{\bf K}{\bf Q}}$ is the effective Hamiltonian providing the energy $\hbar\Delta \xi_{{\bf K},{\bf Q}}=\hbar\xi_{{\bf K}+{\bf Q}}-\hbar\xi_{K}$ of the electron-hole pairs created across the Fermi sea. The electron energy (relative to the Fermi energy) is denoted as $\hbar \xi_{K}$. For simplicity, we assume a 3D spherical Fermi surface providing $\xi_{K}=\hbar (K^{2}-K^{2}_{\rm F})/(2 m)$. The free phonon Hamiltonian reads $H_{\rm pn}=\sum_{{\bf Q},\alpha}\hbar\omega_{0}B^{\dagger}_{{\bf Q}\alpha}B_{{\bf Q}\alpha}$, and the interaction term $H_{\rm P^{2}}$ is proportional to the square polarization field:
\begin{align}
H_{\rm P^{2}}=\frac{1}{2\epsilon_{0}}\int \!d{\bf R} \, {\bf P}^{2} ({\bf R}).
\label{H_p2}
\end{align} 
This term contains the contributions of both plasmons and phonons $\propto {\bf P}_{\rm pl}^{2}$ and $\propto{\bf P}_{\rm pn}^{2}$, respectively, as well as a direct plasmon-phonon interaction term $\propto {\bf P}_{\rm pl}\cdot {\bf P}_{\rm pn}$. Replacing Eqs.~(\ref{elec_pol32}) and (\ref{pho_pol22}) into Eq.~(\ref{H_p2}) and performing the spatial integration, the matter Hamiltonian is derived as
\begin{align}
H_{\rm mat}&=\sum_{{\bf K},{\bf Q}} \hbar\Delta \xi_{{\bf K},{\bf Q}} P^{\dagger}_{{\bf K}{\bf Q}} P_{{\bf K}{\bf Q}} +\sum_{{\bf K},{\bf K'},{\bf Q}} \hbar\Lambda^{\bf Q}_{{\bf K}{\bf K'}} \left(P_{{\bf -K}{\bf -Q}} + P^{\dagger}_{{\bf K}{\bf Q}}\right)\left(P_{{\bf K'}{\bf Q}} + P^{\dagger}_{{\bf -K'}{\bf -Q}}\right)  \nonumber \\
&+\sum_{{\bf Q},\alpha}\hbar\omega_{0} B^{\dagger}_{{\bf Q}\alpha}B_{{\bf Q}\alpha} + \sum_{{\bf Q},\alpha} \frac{\hbar\nu^{2}_{\rm pn}}{4\omega_{0}} \left(B_{{\bf -Q}\alpha} + B^{\dagger}_{{\bf Q}\alpha}\right) \left(B_{{\bf Q}\alpha} + B^{\dagger}_{{\bf -Q}\alpha}\right) \nonumber \\
&+ \sum_{{\bf K},{\bf Q},\alpha} \hbar \mu^{\alpha}_{{\bf K}{\bf Q}} \left(P_{{\bf -K}{\bf -Q}} + P^{\dagger}_{{\bf K}{\bf Q}}\right) \left(B_{{\bf Q}\alpha} + B^{\dagger}_{{\bf -Q}\alpha}\right),
\label{H_mat_nondiag}
\end{align} 
where we have introduced $\Lambda^{\bf Q}_{{\bf K}{\bf K'}}=\frac{1}{2\epsilon_{0}\hbar V}{\bf d}_{{\bf K}{\bf Q}} \cdot {\bf d}_{{\bf K'}{\bf Q}}$ and $\mu^{\alpha}_{{\bf K}{\bf Q}}=\frac{\nu_{\rm pn}}{\sqrt{2\epsilon_{0}\hbar \omega_{0} V}} {\bf d}_{{\bf K}{\bf Q}} \cdot {\bf u}_{\alpha}$, with $\nu_{\rm pn}=\sqrt{\frac{\mathcal{Z}^{2}N}{M\epsilon_{0} a^{3}}}$ the ion plasma frequency. The Hamiltonian Eq.~(\ref{H_mat_nondiag}) can be diagonalized using the Hopfield-Bogoliubov method~\cite{hopfield}, by introducing collective eigenmodes in the form
\begin{align}
\Pi_{{\bf Q}j}=\sum_{\bf K} \left(p_{{\bf K}{\bf Q}j} P_{{\bf K}{\bf Q}}+ \widetilde{p}_{{\bf K}{\bf Q}j} P^{\dagger}_{{\bf -K}{\bf -Q}} \right) + \sum_{\alpha} \left(b_{{\bf Q}\alpha j} B_{{\bf Q}\alpha}+ \widetilde{b}_{{\bf Q}\alpha j} B^{\dagger}_{{\bf -Q}\alpha}\right).
\label{coll_mode}
\end{align}
The condition for these modes to diagonalize the Hamiltonian Eq.~(\ref{H_mat_nondiag}) is then $[\Pi_{{\bf Q}j},H_{\rm mat}]=\hbar \widetilde{\omega}_{j}\Pi_{{\bf Q}j}$. Computing this commutator and introducing the vector
\begin{align}
{\bf Z}_{{\bf Q}j}=\sum_{\bf K} \frac{{\bf d}_{{\bf K}{\bf Q}}}{2\epsilon_{0}\hbar V} \left( p_{{\bf K}{\bf Q}j} - \widetilde{p}_{{\bf K}{\bf Q}j} \right) + \sum_{\alpha} \frac{\nu_{\rm pn}{\bf u}_{\alpha}}{\sqrt{8\epsilon_{0}\hbar V \omega_{0}}} \left( b_{{\bf K}{\bf Q}j} - \widetilde{b}_{{\bf K}{\bf Q}j} \right),
\label{main_guy}
\end{align}
we obtain the transformation coefficients entering Eq.~(\ref{coll_mode}):
\begin{alignat}{2}
p_{{\bf K}{\bf Q}j}&=\frac{-2{\bf d}_{{\bf K}{\bf Q}} \cdot {\bf Z}_{{\bf Q}j}}{\widetilde{\omega}_{j}-\Delta \xi_{{\bf K},{\bf Q}}} &\qquad \widetilde{p}_{{\bf K}{\bf Q}j}&=\frac{-2{\bf d}_{{\bf K}{\bf Q}} \cdot {\bf Z}_{{\bf Q}j}}{\widetilde{\omega}_{j}+\Delta \xi_{{\bf K},{\bf Q}}} \nonumber \\
b_{{\bf Q}\alpha j}&=\nu_{\rm pn} \sqrt{\frac{2\epsilon_{0}\hbar V}{\omega_{0}}} \frac{{\bf u}_{\alpha}\cdot {\bf Z}_{{\bf Q}j}}{\widetilde{\omega}_{j}-\omega_{0}} &\qquad \widetilde{b}_{{\bf Q}\alpha j}&=\nu_{\rm pn} \sqrt{\frac{2\epsilon_{0}\hbar V}{\omega_{0}}} \frac{{\bf u}_{\alpha}\cdot {\bf Z}_{{\bf Q}j}}{\widetilde{\omega}_{j}+\omega_{0}}.
\label{coeff_transf}
\end{alignat}
Introducing the orthogonal basis $({\bf Q}/Q,{\bf u}_{{\bf Q}1},{\bf u}_{{\bf Q}2})$, where ${\bf u}_{{\bf Q}p}$ ($p=1,2$) denote the two transverse polarization vectors, ${\bf Z}_{{\bf Q}j}$ can be decomposed into its longitudinal and transverse components. Solving for the transverse components, one can replace ${\bf Z}_{{\bf Q}j}=Z_{{\bf Q}j} {\bf u}_{{\bf Q}p}$ in Eq.~(\ref{coeff_transf}), and Eq.~(\ref{main_guy}) turns out to be equivalent to $\epsilon_{\rm cr} (\widetilde{\omega}_{j}){\bf Z}_{{\bf Q}j}=0$, where
\begin{align}
\epsilon_{\rm cr} (\omega)=1 - \frac{\omega^{2}_{\rm pl}}{\omega^{2}}- \frac{\nu^{2}_{\rm pn}}{\omega^{2}-\omega^{2}_{0}}
\label{eps_suppl}
\end{align}
represents the transverse dielectric function of the crystal. To obtain this result, we have used the definition of the electronic plasma frequency $\omega_{\rm pl}=\sqrt{\frac{\rho e^{2}}{m\epsilon_{0}}}$ with $\rho$ the electron density, $\sum_{\alpha} {\bf u}_{\alpha} \cdot {\bf u}_{{\bf Q},p} =1$, as well as the identity
\begin{align}
\frac{1}{\hbar \epsilon_{0} V} \sum_{\bf K} \frac{2 \vert {\bf d}_{{\bf K}{\bf Q}} \cdot {\bf u}_{{\bf Q}p} \vert^{2} \Delta \xi_{{\bf K},{\bf Q}}}{\omega^{2}-\Delta\xi^{2}_{{\bf K},{\bf Q}}} \sim \frac{\omega^{2}_{\rm pl}}{\omega^{2}},
\label{ident}
\end{align}
in the so-called dynamical long-wavelength limit $Q \ll K_{\rm F}$, $\omega \gg v_{\rm F}Q$ ($v_{\rm F}=\hbar K_{\rm F}/m$ is the Fermi velocity). Note that in this regime, one can use the relation
\begin{align*}
\vert {\bf d}_{{\bf K}{\bf Q}} \cdot {\bf u}_{{\bf Q}p} \vert^{2} = \frac{4 e^{2} n_{\bf K} \left(1- n_{{\bf K}+{\bf Q}}\right) \left[\left(Q K\right)^{2}- \left({\bf K}\cdot {\bf Q}\right)^{2} \right]}{Q^{2} \left(Q^{2}+ 2{\bf K}\cdot {\bf Q} \right)^{2}} \sim \frac{e^{2} n_{\bf K} \left(1- n_{{\bf K}+{\bf Q}}\right)}{Q^{2}}
\end{align*}
to put Eq.~(\ref{ident}) in the form of the well-known Lindhard function, which describes the longitudinal response of a free electron gas~\cite{Lindhard}. The two eigenvalues $\widetilde{\omega}_{j}$ can be determined by solving the equation $\epsilon_{\rm cr} (\widetilde{\omega}_{j})=0$, which provides the solutions given by Eq.~(1) of the main text. As detailed in Ref.~[\onlinecite{todorovY}], the vector ${\bf Z}_{{\bf Q}j}$ can be decomposed as ${\bf Z}_{{\bf Q}j}=\mathcal{N}_{j} \mathcal{T}_{j} {\bf u}_{{\bf Q}p}$ when solving for the transverse modes, where $\mathcal{T}_{j}$ are determined by computing the residues of the inverse dielectric function
\begin{align*}
\frac{1}{\epsilon_{\rm cr} (\omega)} = 1+ \sum_{j=1,2}\frac{\mathcal{T}^{2}_{j}}{\omega^{2}-\widetilde{\omega}^{2}_{j}},
\end{align*} 
and are derived as
\begin{align}
\mathcal{T}^{2}_{2} = \frac{\omega^{2}_{\rm pl}(\widetilde{\omega}^{2}_{2}-\omega^{2}_{0})+ \nu^{2}_{\rm pn} \widetilde{\omega}^{2}_{2}}{\widetilde{\omega}^{2}_{2}-\widetilde{\omega}^{2}_{1}} \qquad  \mathcal{T}^{2}_{1} = \frac{\omega^{2}_{\rm pl}(\widetilde{\omega}^{2}_{1}-\omega^{2}_{0})+ \nu^{2}_{\rm pn} \widetilde{\omega}^{2}_{1}}{\widetilde{\omega}^{2}_{1}-\widetilde{\omega}^{2}_{2}}. 
\label{oscill}
\end{align} 
The constant $\mathcal{N}_{j}$ can be found by using the normalization condition
\begin{align*}
 \sum_{\bf K} \left(\vert p_{{\bf K}{\bf Q}j} \vert^{2}-\vert \widetilde{p}_{{\bf K}{\bf Q}j} \vert^{2} \right) + \sum_{\alpha}\left(\vert b_{{\bf Q}\alpha j} \vert^{2}-\vert \widetilde{b}_{{\bf Q}\alpha j} \vert^{2} \right) =1,
\end{align*} 
stemming from the commutation relation $[\Pi_{{\bf Q}j},\Pi^{\dagger}_{{\bf Q'}j'}]=\delta_{{\bf Q},{\bf Q'}} \delta_{j,j'}$. Using Eq.~(\ref{coeff_transf}) with ${\bf Z}_{{\bf Q}j}=\mathcal{N}_{j} \mathcal{T}_{j} {\bf u}_{{\bf Q}p}$, as well as the identity
\begin{align*}
\frac{1}{\hbar \epsilon_{0} V} \sum_{\bf K} \frac{2 \vert {\bf d}_{{\bf K}{\bf Q}} \cdot {\bf u}_{{\bf Q}p} \vert^{2} \Delta \xi_{{\bf K},{\bf Q}}}{\left(\omega^{2}-\Delta\xi^{2}_{{\bf K},{\bf Q}}\right)^{2}} \sim \frac{\omega^{2}_{\rm pl}}{\omega^{4}},
\end{align*}
which can be derived similarly as Eq.~(\ref{ident}) in the dynamical long-wavelength limit, we obtain
\begin{align}
\frac{1}{\mathcal{N}^{2}_{j}} = 8\hbar \epsilon_{0}V \widetilde{\omega}_{j} \mathcal{T}^{2}_{j} \left(\frac{\omega^{2}_{\rm pl}}{\widetilde{\omega}_{j}^{4}}+\frac{\nu^{2}_{\rm pn}}{(\widetilde{\omega}_{j}^{2}-\omega^{2}_{0})^{2}} \right).
\label{const}
\end{align}
Finally, we use Eqs.~(\ref{elec_pol32}), (\ref{pho_pol22}), (\ref{coll_mode}), (\ref{coeff_transf}), the decomposition ${\bf Z}_{{\bf Q}j}=\mathcal{N}_{j} \mathcal{T}_{j} {\bf u}_{{\bf Q}p}$, as well as the identities
\begin{align*}
\sum_{j}\frac{\mathcal{T}^{2}_{j}}{\widetilde{\omega}_{j}^{2}-\Delta \xi^{2}_{{\bf K},{\bf Q}}} =1 \qquad \sum_{j}\frac{\mathcal{T}^{2}_{j}}{\widetilde{\omega}_{j}^{2}-\omega^{2}_{0}} =1,
\end{align*}
to write the total (transverse) polarization field as
\begin{align}
{\bf P} ({\bf R}) = \sum_{{\bf Q},j,p} \frac{\mathcal{T}_{j}}{4 \mathcal{N}_{j} \widetilde{\omega}_{j} V} \left(\Pi_{{\bf -Q}j p} + \Pi^{\dagger}_{{\bf Q}j p}\right) e^{-i {\bf Q}\cdot {\bf R}} {\bf u}_{{\bf Q}p}.
\label{total_interm}
\end{align}
Since the crystal is assumed to be isotropic, one can choose the transverse polarization vectors ${\bf u}_{{\bf Q}p}$ to coincide with ${\bf u}_{\alpha}$ ($\alpha=z,\parallelsum$), and using Eqs.~(\ref{oscill}) and (\ref{const}), the polarization field Eq.~(\ref{total_interm}) coincides with Eq.~(\ref{pola_newbasis}). The matter Hamiltonian in the new basis takes the simple form $H_{\rm mat}= \sum_{{\bf Q},\alpha,j} \hbar\widetilde{\omega}_{j} \Pi^{\dagger}_{{\bf Q}j\alpha} \Pi_{{\bf Q}j\alpha}$. Importantly, the full matter Hamiltonian also contains the contribution of the electronic dark modes (electron-hole continuum) associated to individual electrons. Since the weight of a collective mode $\Pi_{{\bf Q}j\alpha}$ on the individual electronic states goes to zero as the number of electrons goes to infinity (which is assumed here), this contribution can be written as $\sum_{\bf K} \hbar \xi_{\bf K} c^{\dagger}_{\bf K}c_{\bf K}$, and is included in the coupling Hamiltonian $H_{\rm el-pn}$ between the electronic dark modes and phonons derived in the last section.

\subsection{Coupling to photons: Surface plasmon-phonon polaritons} 
\label{tot_Hh}

In this section, we provide the detailed derivation of the photon Hamiltonian $H_{\rm pt}$ and the light-matter coupling term $H_{\rm mat-pt}$ entering the polariton Hamiltonian $H_{\rm pol}$. We then explain the self-consistent algorithm used to diagonalize $H_{\rm pol}$ and to find the surface polariton modes of the system. The phonon, plasmon, and photon weights of the polaritons are defined at the end of the section. The displacement and magnetic fields ${\bf D}$ and ${\bf H}$ can be found by generalizing Todorov's approach~\cite{todorovY} to the case of a double interface, writing them as a superposition of the fields generated by each interfaces $m=1,2$: ${\bf D} ({\bf R}) = \sum_{{\bf q},m} \sqrt{\frac{4\epsilon_{0}\hbar c}{S}} e^{i {\bf q}\cdot {\bf r}} {\bf u}_{{\bf q}m} (z) D_{{\bf q}m}$, and ${\bf H} ({\bf R}) = \sum_{{\bf q},m} w_{q} \sqrt{\frac{4\epsilon_{0}\hbar c}{S}} e^{i {\bf q}\cdot {\bf r}} {\bf v}_{{\bf q}m} (z) H_{{\bf q}m}$. Here, $S$ denotes the surface of the crystal, and $w_{q}$ the frequency of the surface polariton modes which are still undetermined at this point. Denoting by ${\bf u}_{\perp}$ the in-plane unit vector perpendicular to both ${\bf u}_{z}$ and ${\bf u}_{\parallelsum}$, and $\theta$ the heaviside function, the mode functions read
\begin{align*}
{\bf u}_{{\bf q}1} (z) &= \left[-\gamma_{\rm d}\theta (-z) e^{\gamma_{\rm d}z}+ \gamma_{\rm cr} \theta (z)\theta (\ell-z) e^{-\gamma_{\rm cr} z} + \gamma_{\rm d} \theta (z-\ell) e^{-\gamma_{\rm cr} \ell} e^{-\gamma_{\rm d} (z-\ell)} \right] {\bf u}_{\parallelsum} \nonumber \\ 
&+ i q \left[\theta (-z) e^{\gamma_{\rm d}z}+ \theta (z)\theta (\ell-z) e^{-\gamma_{\rm cr} z} + \theta (z-\ell) e^{-\gamma_{\rm cr} \ell} e^{-\gamma_{\rm d} (z-\ell)} \right] {\bf u}_{z} \nonumber \\ {\bf u}_{{\bf q}2} (z) &= \left[-\gamma_{\rm d}\theta (-z) e^{-\gamma_{\rm cr} \ell} e^{\gamma_{\rm d}z}- \gamma_{\rm cr} \theta (z)\theta (\ell-z) e^{\gamma_{\rm cr} (z-\ell)} + \gamma_{\rm d} \theta (z-\ell) e^{-\gamma_{\rm d} (z-\ell)} \right] {\bf u}_{\parallelsum} \nonumber \\ 
&+ i q \left[\theta (-z) e^{-\gamma_{\rm cr} \ell} e^{\gamma_{\rm d}z}+\theta (z)\theta (\ell-z) e^{\gamma_{\rm cr} (z-\ell)} + \theta (z-\ell) e^{-\gamma_{\rm d} (z-\ell)}\right] {\bf u}_{z},
\end{align*} 
and
\begin{align*}
{\bf v}_{{\bf q}1} (z) &= \left[\theta (-z) e^{\gamma_{\rm d}z}+ \theta (z)\theta (\ell-z) e^{-\gamma_{\rm cr} z} + \theta (z-\ell) e^{-\gamma_{\rm cr} \ell} e^{-\gamma_{\rm d} (z-\ell)} \right] {\bf u}_{\perp} \nonumber \\
{\bf v}_{{\bf q}2} (z) &= \left[\theta (-z) e^{-\gamma_{\rm cr} \ell} e^{\gamma_{\rm d}z}+\theta (z)\theta (\ell-z) e^{\gamma_{\rm cr} (z-\ell)} + \theta (z-\ell) e^{-\gamma_{\rm d} (z-\ell)}\right]{\bf u}_{\perp}.
\end{align*} 

Replacing these expressions into $H_{\rm pt} = \frac{1}{2\epsilon_{0}}\int \!d{\bf R} \, {\bf D}^{2}({\bf R}) + \frac{1}{2\epsilon_{0}c^{2}}\int \!d{\bf R}\, {\bf H}^{2}({\bf R})$ and performing the integrations, the free photon Hamiltonian is derived as
\begin{align}
H_{\rm pt}= \hbar c\sum_{{\bf q},m,m'} \left(\mathcal{A}^{m m'}_{q} D_{{\bf q}m} D_{{\bf -q}m'}+ \mathcal{B}^{m m'}_{q} H_{{\bf q}m} H_{{\bf -q}m'} \right), 
\label{hemm}
\end{align} 
with the overlap matrix elements
\begin{align*}
\mathcal{A}^{m m'}_{q}&= \frac{q^{2}+\gamma^{2}_{\rm cr}}{\gamma_{\rm cr}} \left(1-e^{-2 \gamma_{\rm cr}\ell}\right) + \frac{q^{2}+\gamma^{2}_{\rm d}}{\gamma_{\rm d}} \left(1+e^{-2 \gamma_{\rm cr}\ell}\right) \nonumber \\
\mathcal{B}^{m m'}_{q}&=\frac{w^{2}_{q}}{c^{2}}\left(\frac{1-e^{-2 \gamma_{\rm cr}\ell}}{\gamma_{\rm cr}} + \frac{1+e^{-2 \gamma_{\rm cr}\ell}}{\gamma_{\rm d}}\right)
\end{align*}
for $m=m'$, and
\begin{align*}
\mathcal{A}^{m m'}_{q}&=2 e^{-2 \gamma_{\rm cr}\ell} \left[ \left(q^{2}-\gamma^{2}_{\rm cr} \right)\ell + \frac{q^{2}+\gamma^{2}_{\rm d}}{\gamma_{\rm d}} \right] \nonumber \\
\mathcal{B}^{m m'}_{q}&=\frac{2 w^{2}_{q}}{c^{2}} e^{-2 \gamma_{\rm cr}\ell} \left(\frac{1+\gamma_{\rm d}\ell}{\gamma_{\rm d}} \right)
\end{align*}
for $m\neq m'$. The Hamiltonian Eq.~(\ref{hemm}) is diagonalized by the transformation $D_{{\bf q}\pm}=\left(D_{{\bf q}2} \pm D_{{\bf q}1} \right)/\sqrt{2}$, where the new field operators $D_{{\bf q}\sigma}$ and $H_{{\bf q}\sigma}$ satisfy the commutation relations $[D^{\dagger}_{{\bf q}\sigma},D^{\dagger}_{{\bf q'}\sigma'}]=[H^{\dagger}_{{\bf q}\sigma},H^{\dagger}_{{\bf q'}\sigma'}]=0$, and $[D_{{\bf q}\sigma},H^{\dagger}_{{\bf q'}\sigma'}]=-i C_{q\sigma} \delta_{{\bf q},{\bf q'}}\delta_{\sigma,\sigma'}$, together with the properties $D^{\dagger}_{{\bf q}\sigma}=D_{{\bf -q}\sigma}$ and $H^{\dagger}_{{\bf q}\sigma}=H_{{\bf -q}\sigma}$. The constant $C_{q\sigma}$ can be determined by using the Maxwell's equation
\begin{align*}
\frac{d}{dt}{\bf D}({\bf R})= \frac{i}{\hbar} [H_{\rm pt},{\bf D}({\bf R})] ={\bm \nabla}\times {\bf H}({\bf R}),
\end{align*}
which provides $C_{q\sigma} = \frac{w_{q\sigma}}{2c\beta_{q\sigma}}$~\cite{todorovY}. One can then write the field ${\bf D} ({\bf R})$ in terms of the new photon eigenmode operators $D_{{\bf q}\pm}$, and replace it in the light-matter coupling Hamiltonian $H_{\rm mat-pt}=-\frac{1}{\epsilon_{0}}\int \!d{\bf R} \, {\bf P} ({\bf R}) \cdot {\bf D} ({\bf R})$. Together with Eq.~(\ref{pola_newbasis}), we obtain
\begin{align}
H_{\rm mat-pt}= -\hbar \omega_{\rm pl} \sqrt{\frac{2 c}{\widetilde{\omega}_{j}\ell}} \sum_{\bf Q} & \bigg\{ D_{{\bf q}+}\left[i q \left(\Pi_{{\bf -Q}j z} + \Pi^{\dagger}_{{\bf Q}j z} \right) \mathcal{F}_{+} (Q) + \gamma_{\rm cr} \left(\Pi_{{\bf -Q}j \parallelsum} + \Pi^{\dagger}_{{\bf Q}j \parallelsum} \right) \mathcal{F}_{-} (Q)\right] \nonumber \\
& \!\!\! -D_{{\bf q}-}\left[i q \left(\Pi_{{\bf -Q}j z} + \Pi^{\dagger}_{{\bf Q}j z} \right) \mathcal{F}_{-} (Q) + \gamma_{\rm cr} \left(\Pi_{{\bf -Q}j \parallelsum} + \Pi^{\dagger}_{{\bf Q}j \parallelsum} \right) \mathcal{F}_{+} (Q)\right] \bigg\}
\label{HH_matel}
\end{align}
after spatial integration. The function $\mathcal{F}_{\pm} (Q)$ stems from the overlap between the displacement and the polarization fields and reads $\mathcal{F}_{\pm} (Q)= \left(\mathcal{F}_{2} (Q) \pm \mathcal{F}_{1} (Q) \right)/\sqrt{2}$ with
\begin{align*}
\mathcal{F}_{1}(Q) =e^{-\gamma_{\rm cr}\ell} \left(\frac{1-e^{-(i q_{z} - \gamma_{\rm cr})\ell}}{i q_{z} - \gamma_{\rm cr}}\right) \qquad \mathcal{F}_{2}(Q) =\frac{1-e^{-(i q_{z} + \gamma_{\rm cr})\ell}}{i q_{z} + \gamma_{\rm cr}}.
\end{align*}
We now introduce the quasi-2D ``bright'' modes $\pi^{\dagger}_{{\bf q}\sigma j}=\sum_{q_{z},\alpha} f_{\alpha \sigma} (Q) \Pi^{\dagger}_{{\bf Q}j\alpha}$ and its hermitian conjugate, which consist of linear superpositions of the 3D plasmon-phonon hybrid modes. The function $f_{\alpha\sigma} (Q)$ reads
\begin{align*}
f_{z\pm} (Q) = i q \mathfrak{N}_{q\pm} \mathcal{F}_{\pm} (Q) \qquad f_{\parallelsum\pm} (Q) = \gamma_{\rm cr} \mathfrak{N}_{q\pm} \mathcal{F}_{\mp} (Q).
\end{align*}
The normalization constant $\mathfrak{N}_{q\sigma}$ is determined by imposing the bosonic commutation relations $[\pi_{{\bf q}\sigma j},\pi^{\dagger}_{{\bf q'}\sigma j}]=\delta_{{\bf q},{\bf q'}}$, which provides $1/\mathfrak{N}^{2}_{q\pm}=\sum_{q_{z}} q^{2}\vert \mathcal{F}_{\pm} (q_{z}) \vert^{2} + \gamma^{2}_{\rm cr} \vert \mathcal{F}_{\mp} (q_{z}) \vert^{2}$. Transforming the summation into an integral $\sum_{q_{z}} \to \ell/(2\pi)\int^{+\infty}_{-\infty} \! dq_{z}$, and using the results
\begin{align}
\int \! d q_{z} \vert \mathcal{F}_{m}(Q) \vert^{2}= \frac{\pi \left(1-e^{-2 \gamma_{\rm cr} \ell} \right)}{\gamma_{\rm cr}} \quad \textrm{for} \quad m=1,2,
\label{raisin_1}
\end{align}
\begin{align}
\int \! d q_{z} \frac{e^{i q_{z} \ell}}{\left(i q_{z} + \gamma_{\rm cr}\right)^{2}} = \int \! d q_{z} \frac{e^{-i q_{z} \ell}}{\left(i q_{z} - \gamma_{\rm cr}\right)^{2}} = 2\pi \ell e^{-\gamma_{\rm cr} \ell},
\label{raisin_12}
\end{align}
and
\begin{align}
\int \! d q_{z} \frac{e^{i q_{z} \ell}}{\left(i q_{z} - \gamma_{\rm cr}\right)^{2}} = \int \! d q_{z} \frac{e^{-i q_{z} \ell}}{\left(i q_{z} + \gamma_{\rm cr}\right)^{2}} = \int \! \frac{d q_{z}}{\left(i q_{z} \pm \gamma_{\rm cr}\right)^{2}} = 0,
\label{raisin_2}
\end{align}
we finally obtain
\begin{align}
\frac{1}{\mathfrak{N}^{2}_{q\sigma}}=\frac{\ell}{2} \left[\frac{q^{2}+\gamma^{2}_{\rm cr}}{\gamma_{\rm cr}} \left(1-e^{-2 \gamma_{\rm cr}\ell}\right) + 2 \ell \sigma e^{-\gamma_{\rm cr}\ell}\left(q^{2}-\gamma^{2}_{\rm cr} \right) \right].
\label{csytes}
\end{align}
Note that the existence of a symmetry plane at $z=\ell/2$ implies that the symmetric and antisymmetric bright modes commute, i.e. $[\pi_{{\bf q}\sigma j},\pi^{\dagger}_{{\bf q'}\sigma' j'}]=\delta_{{\bf q},{\bf q'}}\delta_{\sigma,\sigma'}\delta_{j,j'}$, which can be checked using Eqs.~(\ref{raisin_1}), (\ref{raisin_12}), and (\ref{raisin_2}). On the other hand, the commutation relation with respect to $j$ follows directly from the diagonalization of the matter Hamiltonian presented before. One can then use Eq.~(\ref{csytes}) combined with the definition of the bright modes to write the light-matter coupling Hamiltonian Eq.~(\ref{HH_matel}) in the form given in the main text. Using the unitary transformation $\pi^{\dagger}_{{\bf q}\sigma j}=\sum_{q_{z},\alpha} f_{\alpha \sigma} (Q) \Pi^{\dagger}_{{\bf Q}j\alpha}$ and $\Gamma^{\dagger}_{{\bf q}\sigma j s}=\sum_{q_{z},\alpha} g_{\alpha \sigma s} (Q) \Pi^{\dagger}_{{\bf Q}j\alpha}$, with $[\Gamma_{{\bf q}\sigma j s},\Gamma^{\dagger}_{{\bf q'}\sigma' j' s'}]=\delta_{{\bf q},{\bf q'}}\delta_{\sigma,\sigma'}\delta_{j,j'}\delta_{s,s'}$, $[\pi_{{\bf q}\sigma j},\Gamma^{\dagger}_{{\bf q'}\sigma' j' s}]=[\pi^{\dagger}_{{\bf q}\sigma j},\Gamma^{\dagger}_{{\bf q'}\sigma' j' s}]=0$, and the identities
\begin{align*}
&\sum_{q_{z},\alpha} g_{\alpha \sigma s} (Q) g^{*}_{\alpha \sigma' s'} (Q)=\delta_{s,s'}\delta_{\sigma,\sigma'} \\
&\sum_{q_{z},\alpha} f_{\alpha \sigma} (Q) f^{*}_{\alpha \sigma'} (Q)=\delta_{\sigma,\sigma'} \\
&\sum_{q_{z},\alpha} g_{\alpha \sigma s} (Q) f^{*}_{\alpha \sigma'} (Q)=0  \\ 
&\sum_{\sigma,s} g_{\alpha \sigma s} (Q) g^{*}_{\alpha' \sigma s} (Q') + \sum_{\sigma} f_{\alpha \sigma} (Q) f^{*}_{\alpha' \sigma} (Q') = \delta_{\alpha,\alpha'} \delta_{{\bf Q},{\bf Q'}},
\end{align*}
the matter Hamiltonian can be decomposed as $H_{\rm mat}=\sum_{q,\sigma,j} \hbar \widetilde{\omega}_{j} \pi^{\dagger}_{{\bf q}\sigma j} \pi_{{\bf q}\sigma j} + H_{\rm dark}$, where $H_{\rm dark}=\sum_{q,\sigma,j} \sum_{s=1}^{\infty} \hbar \widetilde{\omega}_{j} \Gamma^{\dagger}_{{\bf q}\sigma j s} \Gamma_{{\bf q}\sigma j s}$ denotes the contribution of the quasi-2D dark modes which do not interact with light. Leaving this contribution aside, the polariton Hamiltonian $H_{\rm pol}$ of the main text can be diagonalized numerically using a Hopfield-Bogoliubov transformation~\cite{hopfield}. We introduce the polariton eigenmodes of the system in the form
\begin{align}
\mathcal{P}_{{\bf q}\sigma \zeta}=\sum_{j=1,2} \left(\mathcal{O}_{q\sigma j \zeta} \pi_{{\bf q}\sigma j} + \mathcal{\widetilde{O}}_{q\sigma j \zeta} \pi^{\dagger}_{{\bf -q}\sigma j} \right) + \mathcal{X}_{q\sigma\zeta} D_{{\bf q}\sigma} + \mathcal{Y}_{q\sigma\zeta} H_{{\bf q}\sigma},
\label{fin_pola}
\end{align}
which satisfy the eigenvalue equation $[\mathcal{P}_{{\bf q}\sigma\zeta},H]=w_{q\sigma\zeta}\mathcal{P}_{{\bf q}\sigma\zeta}$. Computing this commutator, one can show that the polariton frequencies $w_{q\sigma\zeta}$ in each subspace $({\bf q},\sigma)$ are given by the 3 positive eigenvalues of the matrix
\begin{align}
\underline{\mathcal{H}}_{{\bf q}\sigma}=\begin{bmatrix} \widetilde{\omega}_{1} & 0 & 0 & 0 & 0 & i \Omega_{q \sigma 1} C_{q\sigma} \\ 0 & -\widetilde{\omega}_{1} & 0 & 0 & 0 & i \Omega_{q \sigma 1} C_{q\sigma} \\ 0 & 0 & \widetilde{\omega}_{2} & 0 & 0 & i \Omega_{q \sigma 2} C_{q\sigma} \\ 0 & 0 & 0 & -\widetilde{\omega}_{2} & 0 & i \Omega_{q \sigma 2} C_{q\sigma} \\ \Omega_{q \sigma 1} & -\Omega_{q \sigma 1} & \Omega_{q \sigma 2} & -\Omega_{q \sigma 2} & 0 & 2 i c \alpha_{q\sigma} C_{q\sigma} \\ 0 & 0 & 0 & 0 & -2 i c \beta_{q\sigma} C_{q\sigma} & 0.
\end{bmatrix}
\label{big_mat}
\end{align}
Out of these three eigenvalues, only the lowest two correspond to surface modes as located below the light cone. These two surface modes are denoted as lower and upper polaritons with frequencies $w_{q\sigma {\rm LP}}$ and $w_{q\sigma {\rm UP}}$, respectively. In order to compute these frequencies, we use a self-consistent algorithm which consists in starting with a given frequency $w_{q\sigma {\rm LP}}$, then determine the penetration depths in each media from the Helmholtz equation $\epsilon_{n} (w_{q\sigma {\rm LP}})w^{2}_{q\sigma {\rm LP}}/c^{2}=q^{2}-\gamma^{2}_{n}$ with $\epsilon_{\rm d}=1$ and $\epsilon_{\rm cr}$ given by Eq.~(\ref{eps_suppl}), and use the values of $\gamma_{n}$ to compute the parameters entering the matrix Eq.~(\ref{big_mat}). The latter is diagonalized numerically allowing to determine the new $w_{q\sigma {\rm LP}}$, and the whole process is repeated until convergence is reached. This method is applied independently for the symmetric and the antisymmetric modes $\sigma=\pm$.


\noindent The photon, phonon, and plasmon weights of the polaritons modes are directly related to the transformation coefficients entering Eq.~(\ref{fin_pola}). From the polariton normalization
\begin{align*}
\sum_{j=1,2} \left(\vert \mathcal{O}_{q\sigma j\zeta}\vert^{2} - \vert \mathcal{\widetilde{O}}_{q\sigma j\zeta} \vert^{2} \right) -2i C_{q\sigma} \mathcal{X}_{q\sigma\zeta} \mathcal{Y}^{*}_{q\sigma\zeta} =1,
\end{align*}
the photon weight is defined as $W^{\zeta}_{{\rm pt},q\sigma}= -2i C_{q\sigma} \mathcal{X}_{q\sigma\zeta} \mathcal{Y}^{*}_{q\sigma\zeta}$, and we now need to compute the phonon and plasmon weights of the hybrid plasmon-phonon modes $j=1,2$. This can be done by realizing that the frequencies $\widetilde{\omega}_{1}$ and $\widetilde{\omega}_{2}$ given by Eq.~(\ref{polpol}) are the eigenvalues of the effective plasmon-phonon coupling Hamiltonian $H_{\rm eff}= \hbar\omega_{\rm pl} P^{\dagger} P + \hbar\widetilde{\omega}_{0} B^{\dagger} B +  \hbar\Omega \left( P + P^{\dagger} \right) \left( B + B^{\dagger} \right)$, where $P$ and $B$ denote respectively the effective plasmon and phonon bosonic operators, $\Omega=(\nu_{\rm pn}/2) \sqrt{\omega_{\rm pl}/\widetilde{\omega}_{0}}$ is the effective coupling strength, and $\widetilde{\omega}_{0}=\sqrt{\omega^{2}_{0}+\nu^{2}_{\rm pn}}$ the longitudinal phonon frequency. We introduce the normal modes $\mathcal{P}^{0}_{j}= \mathcal{U}_{j} P + \mathcal{\widetilde{U}}_{j} P^{\dagger}  + \mathcal{V}_{j} B + \mathcal{\widetilde{V}}_{j} B^{\dagger}$, which satisfy the eigenvalue equation $[\mathcal{P}^{0}_{j},H_{\rm eff}]=\widetilde{\omega}_{j} \mathcal{P}^{0}_{j}$. The coefficients $\mathcal{U}_{j}$, $\mathcal{\widetilde{U}}_{j}$, $\mathcal{V}_{j}$, and $\mathcal{\widetilde{V}}_{j}$ can then be computed similarly as before by diagonalizing the associated Hopfield matrix. Using the normalization condition $\vert \mathcal{U}_{j} \vert^{2} - \vert \mathcal{\widetilde{U}}_{j} \vert^{2} + \vert \mathcal{V}_{j} \vert^{2} - \vert \mathcal{\widetilde{V}}_{j} \vert^{2}=1$, the phonon and plasmon weights of the polariton $\zeta$ are defined as
\begin{align*}
W^{\zeta}_{{\rm pl},q\sigma}= \sum_{j} \left(\vert \mathcal{U}_{j} \vert^{2} - \vert \mathcal{\widetilde{U}}_{j} \vert^{2} \right) \left(\vert \mathcal{O}_{q\sigma j\zeta}\vert^{2} - \vert \mathcal{\widetilde{O}}_{q\sigma j\zeta} \vert^{2} \right) \nonumber \\
W^{\zeta}_{{\rm pn},q\sigma}= \sum_{j} \left(\vert \mathcal{V}_{j} \vert^{2} - \vert \mathcal{\widetilde{V}}_{j} \vert^{2} \right) \left(\vert \mathcal{O}_{q\sigma j\zeta}\vert^{2} - \vert \mathcal{\widetilde{O}}_{q\sigma j\zeta} \vert^{2} \right),
\end{align*}
and satisfy the sum rule $\sum_{i} W^{\zeta}_{i, q\sigma}=1$ with $i={\rm pt},{\rm pl},{\rm pn}$.

\subsection{Electron-phonon scattering}
\label{elec_phon_cou_sec}

In this section, we derive the expression of the coupling Hamiltonian $H_{\rm el-pn}$ between electronic dark modes and intramolecular phonons~\cite{SCHLUTER19921473,PhysRevB.48.7651,PhysRevB.58.8236}, and show how the electron-phonon coupling parameter $\lambda$ is affected by the phonon dressing. The potential $V({\bf R})=\sum_{i} \widetilde{V} ({\bf R}-{\bf R}_{i})$ generated by the vibrating ions in the crystal is expanded in the vicinity of the equilibrium positions ${\bf R}^{0}_{i}$ as $\widetilde{V} ({\bf R}-{\bf R}_{i}) \approx \widetilde{V} ({\bf R}-{\bf R}^{0}_{i}) - {\bm \nabla}\widetilde{V} ({\bf R}-{\bf R}^{0}_{i}) \cdot {\bf X}_{i}$. The displacement vector ${\bf X}_{i}$ is proportional to the phonon polarization field given by Eq.~(\ref{pho_pol22}), and reads
\begin{align*}
{\bf X}_{i}=\sqrt{\frac{\hbar N a^{3}}{2 M\omega_{0}V}} \sum_{{\bf Q},\alpha} \left(B_{{\bf -Q}\alpha} + B^{\dagger}_{{\bf Q}\alpha}\right) {\bf u}_{\alpha} e^{-i {\bf Q}\cdot {\bf R}_{i}}. 
\end{align*}
Introducing the fermion field $\Psi ({\bf R})= \sum_{{\bf K},i} c_{\bf K} \phi ({\bf R}-{\bf R}_{i}) e^{-i {\bf K}\cdot{\bf R}_{i}}$, where $\phi ({\bf R}-{\bf R}_{i})$ denotes a Wannier function localized on site $i$, the electron-phonon coupling Hamiltonian reads 
\begin{align*}
H_{\rm el-pn}= \sum_{\bf K} \hbar \xi_{K} c^{\dagger}_{\bf K} c_{\bf K} -e \int \!\! d{\bf R} \, \Psi^{\dagger} ({\bf R}) V({\bf R}) \Psi ({\bf R}),  
\end{align*}
where the bare phonon contribution is contained in the polariton Hamiltonian derived in the previous section. Using these expressions and retaining only the diagonal terms evaluated at the same position (intramolecular vibrations), $H_{\rm el-pn}$ takes the form  
\begin{align}
H_{\rm el-pn}= \sum_{\bf K} \hbar \xi_{K} c^{\dagger}_{\bf K} c_{\bf K} + \sum_{{\bf K},{\bf Q},\alpha} \hbar \mathcal{M} c^{\dagger}_{\bf K} c_{{\bf K}-{\bf Q}} \left(B_{{\bf Q}\alpha}+ B^{\dagger}_{{\bf -Q}\alpha}\right),  
\label{elec_phfinal}
\end{align}
with the coupling constant $\mathcal{M}=e V_{00} \sqrt{\frac{N V}{2 \hbar M \omega_{0}a^{3}}}$ independent on the wave vector, and $V_{00}= \int \! d{\bf R}\, \vert \phi ({\bf R})\vert^{2} {\bm \nabla} \widetilde{V} ({\bf R}) \cdot {\bf u}_{\alpha}$. Electron-phonon interactions are often characterised by the dimensionless coupling parameter $\lambda$, which quantifies the electron mass renormalization due to the coupling to phonons~\cite{PhysRevB.6.2577,mahan}. At zero temperature, $\lambda$ reads 
\begin{align}
\lambda= \frac{1}{N_{\rm 3D}}\sum_{\bf K} \delta (\xi_{K}) \Re \left(-\partial_{\omega} \overline{\Sigma}_{\bf K} (\omega) \vert_{\omega=0} \right),
\label{lambda_eq}
\end{align}
where $\Re$ stands for real part, $N_{\rm 3D}=\sum_{\bf K} \delta (\xi_{K})=\frac{V m K_{\rm F}}{2 \pi^2 \hbar}$ is the 3D electron density of states at the Fermi level, and $\partial_{\omega} \overline{\Sigma}_{\bf K} (\omega) \vert_{\omega=0}$ denotes the frequency derivative of the retarded electron self-energy $\overline{\Sigma}_{\bf K} (\omega)$ evaluated at $\omega=0$. An equation of motion analysis~\cite{PhysRev.131.993} of the electron Green's function (GF) $\mathcal{G}_{\bf K} (\tau)=-i \langle c_{\bf K} (\tau) c^{\dagger}_{\bf K} (0) \rangle$ allows to write the electron self-energy as
\begin{align}
\Sigma_{\bf K}(\omega)= \sum_{{\bf Q},\alpha} i \vert \mathcal{M} \vert^{2} \int \!\! \frac{d\omega'}{2\pi} \mathcal{G}_{{\bf K}-{\bf Q}} (\omega+\omega') \mathfrak{B}_{Q\alpha} (\omega'), 
\label{self_eq011}
\end{align}
where $\mathfrak{B}_{Q\alpha} (\omega)=-i\int \! d\tau e^{i\omega \tau} \langle \mathcal{B}_{{\bf Q}\alpha} (\tau)\mathcal{B}_{{\bf -Q}\alpha} (0) \rangle$ is the phonon GF written in the frequency domain, and $\mathcal{B}_{{\bf Q}\alpha}=B_{{\bf Q}\alpha}+B^{\dagger}_{{\bf -Q}\alpha}$. In the absence of phonon-photon coupling ($\nu_{\rm pn}=0$), there is no hybridization between phonons and plasmons, nor is there coupling to photons. Phonons therefore enter the Hamiltonian $H_{\rm pol}$ only in the free contribution $H_{\rm pn}=\sum_{{\bf Q},\alpha}\hbar\omega_{0}B^{\dagger}_{{\bf Q}\alpha}B_{{\bf Q}\alpha}$. In this case, the equation of motion analysis simply provides $\mathfrak{B}_{Q\alpha}(\omega)=2\omega_{0}/(\omega^{2}-\omega^{2}_{0})$. Using this expression together with the non-interacting electron GF $\mathcal{G}^{0}_{\bf K} (\omega)=1/(\omega -\xi_{K})$ in Eq.~(\ref{self_eq011}), the retarded electron self-energy is derived as~\cite{mahan}
\begin{align}
\overline{\Sigma}^{0}_{\bf K}(\omega)= \sum_{{\bf Q},\alpha} \vert \mathcal{M} \vert^{2} \left(\frac{1-n_{{\bf K}-{\bf Q}}}{\omega-\xi_{{\bf K}-{\bf Q}}-\omega_{0}+i0^{+}} + \frac{n_{{\bf K}-{\bf Q}}}{\omega-\xi_{{\bf K}-{\bf Q}}+\omega_{0}+i0^{+}}\right), 
\label{se_anal_rg}
\end{align}
where $0^{+}$ denotes a vanishing positive number. As detailed in the chapter 6 of Ref.~[\onlinecite{mahan}], we keep only the terms $\propto n_{{\bf K}-{\bf Q}}$ in Eq.~(\ref{se_anal_rg}), and calculate the quantity $\partial_{\omega} \overline{\Sigma}_{\bf K} (\omega) \vert_{\omega=0}$. We then use an integration by parts and consider only the term $\propto \int \! d {\bf Q} \frac{\partial n_{{\bf K}-{\bf Q}}}{\partial {\xi_{{\bf K}-{\bf Q}}}}$. We finally obtain the electron-phonon coupling parameter Eqs.~(\ref{lambda_eq}) for $\nu_{\rm pn}=0$:
\begin{align*}
\lambda_{0}= \frac{2 \vert \mathcal{M}\vert^{2}}{N_{\rm 3D}\omega_{0}} \sum_{\bf K} \delta (\xi_{K}) \sum_{\bf Q} \delta (\xi_{{\bf K}-{\bf Q}}) =\frac{2 \vert \mathcal{M} \vert^{2} N_{\rm 3D}}{\omega_{0}}.
\end{align*}
In the presence of phonon-photon coupling ($\nu_{\rm pn}\neq 0$), the phonon dynamics is governed by the Hamiltonian $H_{\rm pol}$, which includes the coupling of phonons to plasmons and photons. Using Eqs.~(\ref{coll_mode}), (\ref{coeff_transf}), (\ref{oscill}), and (\ref{const}), we express the 3D phonon operators $B_{{\bf Q}\alpha}$ and $B^{\dagger}_{{\bf Q}\alpha}$ in terms of the 3D hybrid modes $\Pi_{{\bf Q}\alpha j}$ and $\Pi^{\dagger}_{{\bf Q}\alpha j}$, and then project the latter onto the quasi-2D bright and dark modes defined in the previous section such that the electron-phonon Hamiltonian Eq.~(\ref{elec_phfinal}) takes the form $H_{\rm el-pn}=\sum_{\bf K} \hbar \xi_{K} c^{\dagger}_{\bf K} c_{\bf K}+ H^{\rm (B)}_{{\rm el}-{\rm pn}}+H^{\rm (D)}_{{\rm el}-{\rm pn}}$. The contribution of the bright modes reads
\begin{align*}
H^{\rm (B)}_{{\rm el}-{\rm pn}}= \sum_{{\bf K},{\bf Q}} \sum_{\alpha,\sigma,j} \hbar \mathcal{M} \eta_{j} f^{*}_{\alpha\sigma} (Q) c^{\dagger}_{\bf K} c_{{\bf K}-{\bf Q}} \left(\pi_{{\bf q}\sigma j}+ \pi^{\dagger}_{{\bf -q}\sigma j} \right),
\end{align*}
where $\eta_{j}= \chi_{j} \sqrt{\frac{\omega_{0}/\widetilde{\omega}_{j}}{(\omega_{\rm pl}/\widetilde{\omega}_{j})^{4} + \chi_{j}^{2}}}$ ($\chi_{j}=\frac{\nu_{\rm pn}\omega_{\rm pl}}{\widetilde{\omega}^{2}_{j}-\omega^{2}_{0}}$) is associated to the hybrid modes weights of the phonons. Similarly, the contribution of the dark modes is 
\begin{align*}
H^{\rm (D)}_{{\rm el}-{\rm pn}}=\sum_{{\bf K},{\bf Q}} \sum_{\alpha,\sigma,j,s} \hbar \mathcal{M} \eta_{j} g^{*}_{\alpha\sigma s} (Q) c^{\dagger}_{\bf K} c_{{\bf K}-{\bf Q}} \left(\Gamma_{{\bf q}\sigma j s}+ \Gamma^{\dagger}_{{\bf -q}\sigma j s} \right).
\end{align*}
Using the Hamiltonians $H^{\rm (B)}_{{\rm el}-{\rm pn}}$ and $H^{\rm (D)}_{{\rm el}-{\rm pn}}$, the electron self-energies due to the interaction with bright and dark modes are derived as 
\begin{align}
\Sigma^{\rm (B)}_{\bf K} (\omega)&= \sum_{{\bf Q},\alpha,\sigma,j} i \eta^{2}_{j} \vert \mathcal{M} f_{\alpha\sigma} (Q) \vert^{2} \int \!\! \frac{d\omega'}{2\pi} \mathcal{G}_{{\bf K}-{\bf Q}} (\omega+\omega') \mathfrak{P}_{q\sigma j} (\omega')
\label{self_eq236_11} \\
\Sigma^{\rm (D)}_{\bf K} (\omega)&= \sum_{{\bf Q},\alpha,\sigma,j,s} i \eta^{2}_{j} \vert \mathcal{M} g_{\alpha\sigma s} (Q) \vert^{2} \int \!\! \frac{d\omega'}{2\pi} \mathcal{G}_{{\bf K}-{\bf Q}} (\omega+\omega') \mathfrak{G}_{q\sigma j s} (\omega').
\label{self_eq2387}
\end{align}
Here, $\mathfrak{P}_{q\sigma j} (\omega)$ and $\mathfrak{G}_{q\sigma j s} (\omega)$ denote the bright and dark mode GFs, respectively, defined as $\mathfrak{P}_{q\sigma j} (\omega)=-i\int \! d\tau e^{i\omega \tau} \langle \Upsilon_{{\bf q}\sigma j} (\tau) \Upsilon_{{\bf -q}\sigma j} (0) \rangle$ and $\mathfrak{G}_{q\sigma j s} (\omega)=-i\int \! d\tau e^{i\omega \tau} \langle \Xi_{{\bf q}\sigma j s} (\tau) \Xi_{{\bf -q}\sigma j s} (0) \rangle$, with $\Upsilon_{{\bf q}\sigma j}=\pi_{{\bf q}\sigma j}+\pi^{\dagger}_{{\bf -q}\sigma j}$ and $\Xi_{{\bf q}\sigma j s}=\Gamma_{{\bf q}\sigma j s}+\Gamma^{\dagger}_{{\bf -q}\sigma j s}$. We now decompose the electron-phonon coupling parameter into its bright and dark mode contributions, which are respectively associated to the self-energies Eqs.~(\ref{self_eq236_11}) and (\ref{self_eq2387}): $\lambda=\lambda^{\rm (B)}+\lambda^{\rm (D)}$ for $\nu_{\rm pn}\neq 0$, and $\lambda_{0}=\lambda^{\rm (B)}_{0}+\lambda^{\rm (D)}_{0}$ in the case $\nu_{\rm pn}= 0$. For $\nu_{\rm pn}=0$, the bright and dark mode GFs read $\mathfrak{P}_{q\sigma j} (\omega)=\mathfrak{G}_{q\sigma j s} (\omega)=2\widetilde{\omega}_{j}/(\omega^{2}-\widetilde{\omega}^{2}_{j})$, with $\widetilde{\omega}_{1}=\omega_{0}$, $\widetilde{\omega}_{2}=\omega_{\rm pl}$, $\eta_{1}=1$, and $\eta_{2}=0$. Using these GFs together with $\mathcal{G}^{0}_{\bf K} (\omega)=1/(\omega -\xi_{K})$ in Eqs.~(\ref{self_eq236_11}) and (\ref{self_eq2387}), we proceed as before and find
\begin{align}
\lambda^{\rm (B)}_{0}&=  \frac{2 \vert \mathcal{M}\vert^{2}}{N_{\rm 3D}\omega_{0}} \sum_{\bf K} \delta (\xi_{K}) \sum_{{\bf Q},\alpha,\sigma} \vert f_{\alpha\sigma} (Q) \vert^{2}\delta (\xi_{{\bf K}-{\bf Q}}) \\
\lambda^{\rm (D)}_{0}&=  \frac{2 \vert \mathcal{M}\vert^{2}}{N_{\rm 3D}\omega_{0}} \sum_{\bf K} \delta (\xi_{K}) \sum_{{\bf Q},\alpha,\sigma,s} \vert g_{\alpha\sigma s} (Q) \vert^{2}\delta (\xi_{{\bf K}-{\bf Q}}).
\label{lambda_eq1_BD}
\end{align}
For $\nu_{\rm pn}\neq 0$, the dark modes become hybrid plasmon-phonon modes, but do not interact with photons. Still using the same procedure with the GF $\mathfrak{G}_{q\sigma j s} (\omega)=2\widetilde{\omega}_{j}/(\omega^{2}-\widetilde{\omega}^{2}_{j})$ in Eq.~(\ref{self_eq2387}) and $\mathcal{G}^{0}_{\bf K} (\omega)=1/(\omega -\xi_{K})$, we obtain
\begin{align}
\lambda^{\rm (D)}=  \frac{2 \vert \mathcal{M}\vert^{2}}{N_{\rm 3D}} \sum_{j} \frac{\eta^{2}_{j}}{\widetilde{\omega}_{j}} \sum_{\bf K} \delta (\xi_{K}) \sum_{{\bf Q},\alpha,\sigma,s} \vert g_{\alpha\sigma s} (Q) \vert^{2}\delta (\xi_{{\bf K}-{\bf Q}}).
\label{lambda_eq2_BD}
\end{align}
Interestingly, the sum rule $\sum_{j} \frac{\eta^{2}_{j}}{\widetilde{\omega}_{j}}=\frac{1}{\omega_{0}}$ implies that $\lambda^{\rm (D)}=\lambda^{\rm (D)}_{0}$, which means that the hybridization between plasmons and phonons can not solely lead to a modification of the electron-phonon scattering at the level of the RPA.

We now use the equation of motion theory~\cite{PhysRev.131.993,PhysRevB.97.205303} to calculate the bright mode GF $\mathfrak{P}_{q\sigma j} (\omega)$ for $\nu_{\rm pn} \neq 0$: We start from the Heisenberg equation of motion 
\begin{align*}
\frac{\partial^{2}\Upsilon_{{\bf q}\sigma j} (\tau)}{\partial\tau^{2}}= - [H_{\rm pol},[H_{\rm pol},\Upsilon_{{\bf q}\sigma j}]] (\tau),
\end{align*}
and calculate the double commutator. A vanishing source term $\sum_{{\bf q},\sigma,j} \mathcal{J}_{{\bf q}\sigma j}\Upsilon_{{\bf -q}\sigma j}$ is then added to $H_{\rm pol}$, and we calculate the functional derivative of the obtained equation of motion with respect to $\mathcal{J}_{{\bf q}\sigma j} (0)$. The bright mode GF can be written as
\begin{align*}
\mathfrak{P}_{q\sigma j} (\tau)= -i \frac{\langle \Upsilon_{{\bf q}\sigma j} (\tau)\Upsilon_{{\bf -q}\sigma j} (0) e^{-i \int\! d\tau_{1} H_{\rm pol} (\tau_{1})}\rangle_{0}}{\langle e^{-i \int\! d\tau_{1} H_{\rm pol} (\tau_{1})}\rangle_{0}}= \frac{\delta \langle \Upsilon_{{\bf q}\sigma j} (\tau)\rangle}{\delta \mathcal{J}_{{\bf q}\sigma j} (0)}, 
\end{align*}
where $\langle \cdots \rangle_{0}$ denotes the expectation value in the non-interacting ground state of $H_{\rm pol}$ (without the light-matter coupling term $H_{\rm mat-pt}$). Introducing the photon GF $\mathfrak{D}_{q\sigma} (\tau)=\frac{\delta \langle D_{{\bf q}\sigma} (\tau)\rangle}{\delta \mathcal{J}_{{\bf q}\sigma j} (0)}$, we obtain the following set of equations in the frequency domain:
\begin{align*}
\mathfrak{D}_{q\sigma} (\omega)&=\frac{C_{q\sigma}}{2} \sqrt{\frac{\beta_{q\sigma}}{\alpha_{q\sigma}}} \mathfrak{D}^{0}_{q\sigma} (\omega) \sum_{j} \Omega_{q\sigma j} \mathfrak{P}_{q\sigma j} (\omega) \nonumber \\
\mathfrak{P}_{q\sigma j} (\omega)&=\mathfrak{P}^{0}_{q\sigma j} (\omega)+\Omega_{q\sigma j} \mathfrak{P}^{0}_{q\sigma j} (\omega) \mathfrak{D}_{q\sigma} (\omega),
\end{align*}  
where $\mathfrak{D}^{0}_{q\sigma} (\omega)=\frac{2 W_{q\sigma}}{\omega^{2}-W^{2}_{q\sigma}}$ and $\mathfrak{P}^{0}_{q\sigma j} (\omega)=2\widetilde{\omega}_{j}/(\omega^{2}-\widetilde{\omega}^{2}_{j})$ denote the non-interacting photon and bright mode GFs, and $W_{q\sigma}=2c C_{q\sigma} \sqrt{\alpha_{q\sigma} \beta_{q\sigma}}$ plays the role of the photon frequency. Solving for $\mathfrak{P}_{q\sigma j} (\omega)$, we obtain 
\begin{align}
\mathfrak{P}_{q\sigma j} (\omega)=\frac{\mathfrak{P}^{0}_{q\sigma j} \left(1- \mathcal{S}^{j'j'}_{q\sigma} \mathfrak{P}^{0}_{q\sigma j'} \right)+ \mathfrak{P}^{0}_{q\sigma j}\mathfrak{P}^{0}_{q\sigma j'} \mathcal{S}^{jj'}_{q\sigma}}{\left(1- \mathcal{S}^{jj}_{q\sigma} \mathfrak{P}^{0}_{q\sigma j} \right)\left(1- \mathcal{S}^{j'j'}_{q\sigma} \mathfrak{P}^{0}_{q\sigma j'} \right)-\mathfrak{P}^{0}_{q\sigma j}\mathfrak{P}^{0}_{q\sigma j'} \left(\mathcal{S}^{jj'}_{q\sigma}\right)^{2}} \qquad j'\neq j,
\label{pho_GFGf}
\end{align}
with the photon self-energy $\mathcal{S}^{jj'}_{q\sigma} (\omega)=\frac{2 c C^{2}_{q\sigma} \beta_{q\sigma} \Omega_{q \sigma j} \Omega_{q \sigma j'}}{\left(\omega^{2} - W^{2}_{q\sigma} \right)}$. Note that the polariton frequencies $w_{q\sigma\zeta}$ correspond to the poles of $\mathfrak{P}_{q\sigma j} (\omega)$, and can be determined by solving the associated sixth-order polynomial equation. Proceeding as before, we use the GF Eq.~(\ref{pho_GFGf}) together with $\mathcal{G}^{0}_{\bf K} (\omega)=1/(\omega -\xi_{K})$ in Eq.~(\ref{self_eq236_11}), as well as Eqs.~(\ref{lambda_eq1_BD}) and (\ref{lambda_eq2_BD}), and obtain the relative enhancement of the electron-phonon coupling parameter: 
\begin{align*}
\frac{\Delta\lambda}{\lambda_{0}}\equiv \frac{\lambda-\lambda_{0}}{\lambda_{0}}= \frac{1}{N^{2}_{\rm 3D}} \sum_{\bf K} \delta (\xi_{K}) \sum_{{\bf Q},\alpha,\sigma} \left(\varphi_{q\sigma} - 1 \right) \vert f_{\alpha\sigma} (Q) \vert^{2} \delta (\xi_{{\bf K}-{\bf Q}}),
\end{align*}
where the function $\varphi_{q\sigma \zeta}$ describes the renormalization of the phonon energy due to the coupling to plasmons and photons, and depends on the polariton frequencies as
\begin{align*}
\varphi_{q\sigma} = \sum_{\zeta} \sum_{j,j'}\eta^{2}_{j} \widetilde{\omega}_{j}\omega_{0}\left(1-\delta_{j,j'} \right)\frac{\left(w^{2}_{q\sigma\zeta} - \widetilde{\omega}^{2}_{j'}\right) \left(w^{2}_{q\sigma\zeta} - W^{2}_{q\sigma} \right) + 4 c C^{2}_{q\sigma} \beta_{q\sigma} \widetilde{\omega}_{j'} \Omega_{q \sigma j'} \left(\Omega_{q \sigma j}  - \Omega_{q \sigma j'} \right)}{w^{2}_{q\sigma\zeta} \prod_{\zeta'\neq \zeta} \left(w^{2}_{q\sigma\zeta} - w^{2}_{q\sigma\zeta'} \right)}.
\end{align*} 
One can then transform the summation over ${\bf Q}$ into an integral, and write the relative enhancement in the simple form $\frac{\Delta\lambda}{\lambda_{0}}= \int^{1}_{0} \! dx \,x F(x)$, where $x=Q/(2 K_{\rm F})$, and $F (x)$ is a dimensionless function defined as $F (\widetilde{q},\widetilde{q}_{z})= \sum_{\sigma,\alpha} \left(\varphi_{q\sigma} -1 \right) \vert f_{\alpha\sigma} (Q) \vert^{2}$ with the replacements $\widetilde{q}=q/(2K_{\rm F}) \to x\sqrt{1-x^{2}}$ and $\widetilde{q}_{z}=q_{z}/(2K_{\rm F})\to x^{2}$. 


\subsection{Effective quantum model in the USC regime} 
\label{garcia_ripol_vidal}

\begin{figure}[ht]
\centerline{\includegraphics[width=0.7\columnwidth]{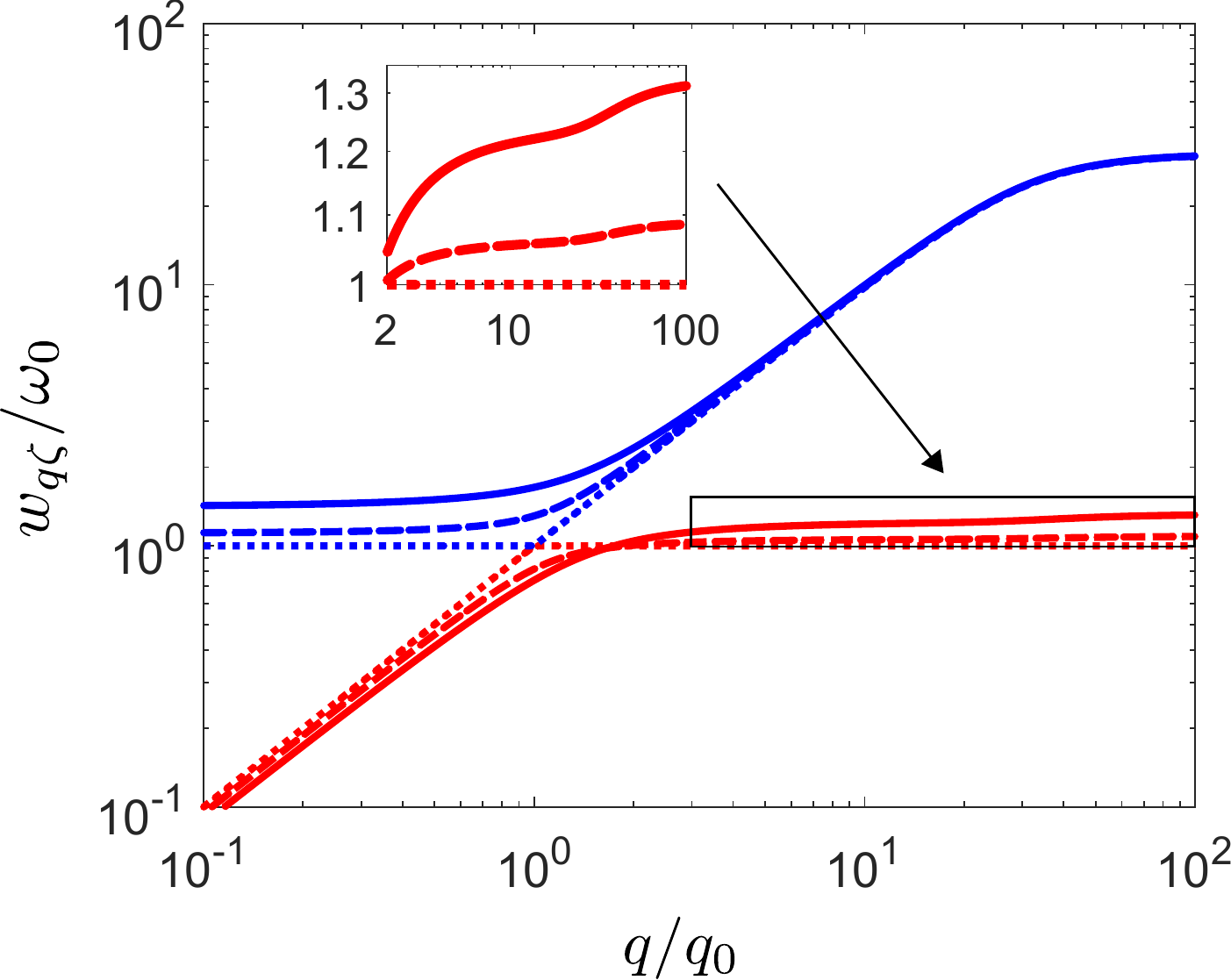}}
\caption{Log-log scale: Normalized polariton frequencies $w_{q\zeta}/\omega_{0}$ versus in-plane wave vector $q/q_{0}$ ($q_{0}=\omega_{0}/c$), for $\nu_{\rm pn}=0.01\omega_{0}$ (dotted lines), $\nu_{\rm pn}=0.5\omega_{0}$ (dashed lines), and $\nu_{\rm pn}=\omega_{0}$ (solid lines). The lower $\zeta={\rm LP}$ and upper $\zeta={\rm UP}$ polaritons are depicted as red and blue lines. The region delimited by the rectangle is magnified in the inset. The plasma frequency is chosen as $\omega_{\rm m}=45 \omega_{0}$, which corresponds to $\hbar \omega_{\rm m}=9 {\rm eV}$ (gold or silver) for $\hbar \omega_{0}=200 {\rm meV}$ (mid-infrared optical phonons).} 
\label{fig_sm}
\end{figure}

In this section, we show that an effective quantum model similar to that of Ref.~[\onlinecite{tudela_2013}] exhibits unphysical behaviors in the ultrastrong coupling regime. We consider a single metal-dielectric interface featuring a plasmon mode with plasma frequency $\omega_{\rm m}$ in the metal region, and a collection of vibrating ions with frequency $\omega_{0}$ forming a lattice embedded in the dielectric region of permittivity $1$ (e.g. air). Phonons are polarized in the directions ${\bf u}_z$ and ${\bf u}_{\parallelsum}$, and both the metal and dielectric regions extend to infinity on each side. In the effective quantum model of Ref.~[\onlinecite{tudela_2013}], the Power-Zienau-Woolley Hamiltonian of the system reads $H_{\rm pol}=H_{\rm spp}+ H_{\rm pn-spp}+H_{\rm pn}+H_{\rm P^{2}}$, where $H_{\rm spp}$ is the effective SPP Hamiltonian, $H_{\rm pn-spp}$ is the SPP-phonon coupling, and $H_{\rm pn}=\sum_{{\bf Q},\alpha}\hbar\omega_{0}B^{\dagger}_{{\bf Q}\alpha}B_{{\bf Q}\alpha}$ is the phonon Hamiltonian. The term proportional to the square polarization reads $H_{\rm P^{2}}=\frac{1}{2\epsilon_{0}}\int \!d{\bf R} \, {\bf P}^{2}_{\rm pn} ({\bf R})$, where the phonon polarization field ${\bf P}_{\rm pn}$ is given by Eq.~(\ref{pho_pol22}). One can show that the overall phonon contribution to the Hamiltonian can be put in the diagonal form $H_{\rm pn}+ H_{\rm P^{2}}= \hbar \widetilde{\omega}_{0} \sum_{{\bf Q},\alpha} \widetilde{B}^{\dagger}_{{\bf Q}\alpha} \widetilde{B}_{{\bf Q}\alpha}$, with the transformation 
\begin{align}
B_{{\bf Q},\alpha}= \frac{\widetilde{\omega}_{0}+\omega_{0}}{2\sqrt{\omega_{0}\widetilde{\omega}_{0}}}\widetilde{B}_{{\bf Q},\alpha} - \frac{\widetilde{\omega}_{0}-\omega_{0}}{2\sqrt{\omega_{0}\widetilde{\omega}_{0}}}\widetilde{B}^{\dagger}_{{\bf -Q},\alpha} \qquad
B^{\dagger}_{{\bf -Q},\alpha}= -\frac{\widetilde{\omega}_{0}-\omega_{0}}{2\sqrt{\omega_{0}\widetilde{\omega}_{0}}}\widetilde{B}_{{\bf Q},\alpha} + \frac{\widetilde{\omega}_{0}+\omega_{0}}{2\sqrt{\omega_{0}\widetilde{\omega}_{0}}}\widetilde{B}^{\dagger}_{{\bf -Q},\alpha}.
\label{bogo_opp}
\end{align}
The longitudinal phonon frequency is defined as $\widetilde{\omega}_{0}=\sqrt{{\omega}^{2}_{0}+\nu^{2}_{\rm pn}}$ with $\nu_{\rm pn}$ the ion plasma frequency. The effective SPP Hamiltonian reads 
\begin{align*}
H_{\rm spp} = \frac{1}{2\epsilon_{0}}\int \!d{\bf R} \, {\bf D}^{2}({\bf R}) + \frac{1}{2\epsilon_{0}c^{2}}\int \!d{\bf R}\, {\bf H}^{2}({\bf R}).
\end{align*}
In contrast to the model presented in the previous sections, here, the displacement and magnetic fields ${\bf D}$ and ${\bf H}$ contain the contribution of the plasmon polarization and \textit{do not involve pure photonic degrees of freedom.} The displacement field can be written as
\begin{align}
{\bf D}({\bf R}) = i \sum_{\bf q} \sqrt{\frac{\epsilon_{0}\hbar w^{0}_{q}}{2 S L_{\rm eff}(w^{0}_{q})}} (a_{\bf q}-a^{\dagger}_{\bf -q}) {\bf U}_{q} e^{i {\bf q}\cdot {\bf r}},
\label{dispEE}
\end{align}
with $S$ an arbitrary (large) surface in the plane, and $a_{\bf q}$ and $a^{\dagger}_{\bf q}$ the annihilation and creation operators of a SPP with frequency $w^{0}_{q}$, which satisfy the bosonic commutation relations $[a_{\bf q},a^{\dagger}_{\bf q'}]=\delta_{{\bf q},{\bf q'}}$. The mode functions read
\begin{align*}
{\bf U}_{q} = \left({\bf u}_{\parallelsum} - \frac{q}{\gamma_{\rm d}} {\bf u}_{z} \right) e^{-\gamma_{\rm d}z},
\end{align*}
and the penetration depths in the metal and dielectric regions $\gamma_{\rm m}$ and $\gamma_{\rm d}$ are related to the SPP frequency via the Helmholtz equation $\gamma_{\rm m}=\sqrt{q^{2}- \epsilon_{\rm m} (\xi_{q}) \left(w^{0}_{q}\right)^{2}/c^{2}}$ and $\gamma_{\rm d}=\sqrt{q^{2}-\left(w^{0}_{q}\right)^{2}/c^{2}}$. The SPP frequency $w^{0}_{q}$ is solution of the equation
\begin{align*}
\omega= q c \sqrt{\frac{\epsilon_{\rm m} (\omega)+1}{\epsilon_{\rm m} (\omega)}},
\end{align*}
with $\epsilon_{\rm m} (\omega)=1 - (\omega_{\rm m}/\omega)^{2}$ the Drude permittivity of the (lossless) metal. According to the SPP quantization scheme detailed in Ref.~[\onlinecite{PhysRevB.82.035411}], the effective quantization length $L_{\rm eff} (\omega)$ is defined as
\begin{align*}
2 L_{\rm eff}(\omega) = \frac{1}{2\gamma_{\rm m}} \left[\left(1+\frac{q^{2}}{\gamma^{2}_{\rm m}} \right) \frac{d \left[\omega \epsilon_{\rm m} (\omega)\right]}{d\omega} + \bigg\vert \frac{\epsilon_{\rm m} (\omega)}{\gamma_{\rm m}} \bigg\vert^{2} \frac{\omega^{2}}{c^{2}} \right] +
 \frac{1}{2\gamma_{\rm d}} \left[\left(1+\frac{q^{2}}{\gamma^{2}_{\rm d}} \right) + \frac{1}{\vert \gamma_{\rm d}\vert^{2}} \frac{\omega^{2}}{c^{2}} \right].
\end{align*}
With these definitions, one can show~\cite{PhysRevB.4.4129,kimMS,PhysRevB.82.035411} that the effective SPP Hamiltonian takes the usual form $H_{\rm spp}=\sum_{\bf q} w^{0}_{q} a^{\dagger}_{\bf q}a_{\bf q}$. Expressing the phonon polarization field Eq.~(\ref{pho_pol22}) in terms of the new operators $\widetilde{B}_{{\bf Q},\alpha}$ and $\widetilde{B}^{\dagger}_{{\bf -Q},\alpha}$ defined by Eq.~(\ref{bogo_opp}), and using the displacement field Eq.~(\ref{dispEE}), the phonon-SPP coupling Hamiltonian $H_{\rm ph-spp}=-\frac{1}{\epsilon_{0}}\int \!d{\bf R} \, {\bf P}_{\rm pn} ({\bf R}) \cdot {\bf D} ({\bf R})$ takes the form
\begin{align*}
H_{\rm ph-spp} = \sum_{\bf q} i \hbar \Omega_{q} \left(b_{\bf -q}+ b^{\dagger}_{\bf q} \right)\left(a_{\bf q} - a^{\dagger}_{\bf -q} \right),
\end{align*}
with the vacuum Rabi frequency
\begin{align*}
\Omega_{q} = \nu_{\rm pn} \sqrt{\frac{w^{0}_{q}}{8 \widetilde{\omega}_{0} L_{\rm eff}(w^{0}_{q}) \gamma_{\rm d}}} \sqrt{1+\frac{q^{2}}{\gamma^{2}_{\rm d}}}.
\end{align*}
The quasi-2D phonon operators are defined as $b_{\bf q} = \mathfrak{N}_{q} \sum_{q_{z}} \frac{(q/\gamma_{\rm d}) \widetilde{B}_{{\bf Q},z}- \widetilde{B}_{{\bf Q},\parallelsum}}{\gamma_{\rm d}- i q_{z}}$, where the normalization $\mathfrak{N}_{q}$ is determined by the bosonic commutation relation $[b_{\bf q},b^{\dagger}_{\bf q}]=1$, which provides $\frac{1}{\mathfrak{N}^{2}_{q}}=\sum_{q_{z}} \frac{1+(q/\gamma_{\rm d})^{2}}{\gamma^{2}_{\rm d}+ q^{2}_{z}}$. Gathering the different contributions, the polariton Hamiltonian reads
\begin{align*}
H_{\rm pol} = \sum_{\bf q} \hbar w^{0}_{q} a^{\dagger}_{\bf q}a_{\bf q} + \sum_{\bf q} i \hbar \Omega_{q} \left(b_{\bf -q}+ b^{\dagger}_{\bf q} \right)\left(a_{\bf q} - a^{\dagger}_{\bf -q} \right) + \hbar \widetilde{\omega}_{0} \sum_{\bf q} b^{\dagger}_{\bf q}b_{\bf q},
\end{align*}
and can be diagonalized using a Hopfield-Bogoliubov transformation. The two resulting polaritons modes are represented on Fig.~\ref{fig_sm} for $\nu_{\rm pn}=0.01\omega_{0}$ (dotted lines), $\nu_{\rm pn}=0.5\omega_{0}$ (dashed lines), and $\nu_{\rm pn}=\omega_{0}$ (solid lines). The lower $\zeta={\rm LP}$ and upper $\zeta={\rm UP}$ polaritons with frequency $w_{q\zeta}$ are depicted as red and blue lines, respectively. While the limiting behaviors of the LP $w_{q{\rm LP}}\sim q c$ for $q \to 0$ and the UP $w_{q{\rm UP}}\to \omega_{\rm m}/\sqrt{2}$ for $q \to \infty$ are properly described by the model, $w_{q{\rm UP}}$ for $q \to 0$ and $w_{q{\rm LP}}$ for $q \to \infty$ feature unphysical behaviors in the ultrastrong coupling regime, i.e. when $\nu_{\rm pn}$ is a non-negligible fraction of $\omega_{0}$. Indeed, $w_{q{\rm LP}}$ exhibits irregularities in the regime of large detunings $q \gg q_{0}$ (see inset), and becomes larger than $\omega_{0}$ while increasing the coupling strength $\nu_{\rm pn}$. We conclude that the effective quantum model based on quantized SPPs is well suited as long as the coupling strength does not exceed a few percent of the associated transition frequency, and therefore fails in the ultrastrong coupling regime.

\pagebreak

\bibliography{supra_light}

\end{document}